\documentclass[usegraphicx,usenatbib,usedcolumn]{mn2e}
\usepackage{longtable}
\usepackage{mathrsfs}
\title[The common envelope ejection efficiency]{Is the common envelope ejection efficiency a function of the binary parameters?}

\author[P.~J.~Davis et al.]{P.~J.~Davis,$^1$ U.~Kolb,$^2$ C.~Knigge$^3$
\\ $^1$Universit\'{e} Libre de Bruxelles, Institut d'Astronomie et d'Astrophysique, 
 Boulevard du Triomphe, B-1050 Brussels, Belgium \\ 
 $^2$The Open University, Department of
 Physics and Astronomy, Walton Hall, Milton Keynes MK7 6AA \\
 $^3$University of Southampton, School of Physics and Astronomy,
 Highfield, Southampton, SO17 1BJ}
\begin{document}

\maketitle
\begin{abstract}
  We reconstruct the common envelope (CE) phase for the current sample of
  observed white dwarf-main sequence post-common envelope binaries
  (PCEBs). We apply multi-regression analysis in order to investigate
  whether correlations exist between the CE ejection efficiencies,
  $\alpha_{\rmn{CE}}$ inferred from the sample, and the binary parameters:
  white dwarf mass, secondary mass, orbital period at the point the CE
  commences, or the orbital period immediately after the CE phase. We do
  this with and without consideration for the internal energy of the
  progenitor primary giants' envelope. Our fits should pave the first steps
  towards an observationally motivated recipe for calculating
  $\alpha_{\rmn{CE}}$ using the binary parameters at the start of the CE
  phase, which will be useful for population synthesis calculations or
  models of compact binary evolution. If we do consider the internal energy
  of the giants' envelope, we find a statistically significant correlation
  between $\alpha_{\rmn{CE}}$ and the white dwarf mass. If we do not, a
  correlation is found between $\alpha_{\rmn{CE}}$ and the orbital period
  at the point the CE phase commences. Furthermore, if the internal energy
  of the progenitor primary envelope is taken into account, then the CE
  ejection efficiencies are within the canonical range
  $0<\alpha_{\rmn{CE}}\leq{1}$, although PCEBs with brown dwarf secondaries
  still require $\alpha_{\rmn{CE}}\ga{1}$.

\end{abstract}

\begin{keywords}
binaries: close -- stars: evolution -- methods: statistical -- white dwarfs
\end{keywords}

\section{Introduction}
\label{section:intro}

The common envelope (CE) phase is a key formation process of all
compact binary systems, such as cataclysmic variables (CVs) and double
white dwarf binaries. Such systems typically have orbital separations
of a few solar radii, and yet require orbital separations of between
approximately 10 and 1000 R$_{\odot}$ to accommodate the giant
progenitor primary star.

\citet{Paczynski1976} first suggested that the CE phase is responsible for
removing large amounts of orbital energy and angular momentum from the
progenitor system, causing a significant reduction in the orbital
separation between the two stellar components \citep[for reviews
see][]{Iben1993,Webbink2008}.

In the pre-CE phase of evolution, the initially more massive stellar
component (which we henceforth denote as the primary) evolves off the main
sequence first. Depending on the orbital separation of the binary, the
primary will fill its Roche lobe on either the red giant or asymptotic
giant branch (AGB) and initiate mass transfer. If the giant primary
possesses a deep convective envelope [i.e. the convective envelope has a
mass of more than approximately 50 per cent of the giant's mass;
\citet{Hjellming1987}], the giant will expand in response to rapid mass
loss. As a result, the giant's radius expands relative to its Roche lobe
radius, increasing the mass transfer rate.  As a consequence of this
run-away situation, mass transfer commences on a dynamical timescale. The
companion main-sequence star (henceforth the secondary) cannot incorporate
this material into its structure quickly enough and therefore expands to
fill its own Roche lobe. The envelope eventually engulfs both the core of
the primary and the main-sequence secondary.

During the spiral-in phase, if enough orbital energy is imparted to
the CE before the stellar components merge, then the CE may be ejected
from the system leaving the core of the primary (now the white dwarf)
and the secondary star at a greatly reduced separation. We term such
systems post-common envelope binaries (PCEBs). In the present study,
we consider PCEBs which have a white dwarf primary component, and a
main sequence secondary companion.

Modelling the CE phase presents a major computational challenge, and can
only be adequately accomplished via three-dimensional (3D) hydrodynamical
simulations \citep[e.g.][]{Sandquist2000}. Even this approach, however,
still cannot cope with the large dynamic range of time and length scales
involved during the CE phase.

Clearly, such hydrodynamical simulations are too computationally intensive
to be included in full binary or population synthesis codes. Instead, such
codes resort to parameterisations of the CE phase. One such parametrization
describes the CE phase in terms of a simple energy budget argument. A
fraction $\alpha_{\rmn{CE}}$ of the orbital energy released as the binary
orbit tightens, $\Delta{E}_{\rmn{orb}}$, is available to unbind the giant's
envelope from the core. Hence, if the change in the envelope's binding
energy is $\Delta{E}_{\rmn{bind}}$, we have
\citep[e.g.][]{deKool1992,deKool1993,Willems2004}
\begin{equation}
\Delta{E}_{\rmn{bind}}=\alpha_{\rmn{CE}}\Delta{E}_{\rmn{orb}}.
\label{Ebind_1}
\end{equation}
The efficiency $\alpha_{\rmn{CE}}$ is a free parameter with values
$0<\alpha_{\rmn{CE}}\leq{1}$ (although values of $\alpha_{\rmn{CE}}>1$
are discussed). However, the value of $\alpha_{\rmn{CE}}$ is poorly
constrained as a result of our equally poor understanding of the
physics underlying the CE phase.

An alternative formulation in terms of the angular momentum budget of
the binary was suggested by \citet{Nelemans2000} from their
investigation into the formation of double white dwarf binaries, which
they assumed occurred from two CE phases. They suggested that the
relative change in the binary's total angular momentum, $\Delta{J}/J$,
and the relative change in the binary's total mass, $\Delta{M}/M$,
during the CE phase are related by
\begin{equation}
\frac{\Delta{J}}{J}=\gamma_{\rmn{CE}}\frac{\Delta{M}}{M},
\label{AM_budget}
\end{equation}
where $\gamma_{\rmn{CE}}$ is the specific angular momentum removed
from the binary by the ejected CE in units of the binary's initial
specific angular momentum. \citet{Nelemans2005} found that a value of
$1.5\leq{\gamma_{\rmn{CE}}}\leq{1.75}$ could account for all observed
PCEBs.

This alternative description was prompted by the fact that for double white
dwarf binaries a value of $\alpha_{\rmn{CE}}<0$ was needed to describe the
first CE phase, which is clearly unphysical. This indicates that the
orbital energy increases during the first phase of mass transfer, with a
corresponding increase in the orbital separation. In light of this,
\citet{Webbink2008} suggested that the first phase of mass transfer is
quasi-conservative, and does not give rise to a CE phase.

However, the angular momentum budget approach predicts PCEB local
space densities that are approximately a factor of ten larger than
observed estimates \citep{Davis2010}. It also predicts that the number
of systems increases towards longer orbital periods, with a maximum
number of systems at approximately 1000 d
\citep{maxted07,Davis2010}. This is contrary to observations, which
show that there is a maximum number of PCEBs with orbital periods of
about 1 d \citep{Mansergas2008}. Furthermore, the angular momentum
budget approach has far less predictive power than the energy budget
description; a tightly constrained value of $\gamma_{\rmn{CE}}$ cannot
constrain the parameters of the possible progenitors of observed
systems, while the energy budget approach is more promising in this
respect \citep{Beer2007,Zorotovic2010}.

The energy budget approach therefore appears to be the more favourable
approach. However, a few major unresolved issues have yet to be
overcome if we are to make significant progress in understanding and
modelling the CE phase.

The first issue is that we require $\alpha_{\rmn{CE}}>1$ to account
for some observed systems such as the double white dwarf binary PG
1115+166 \citep{Maxted2002a} and the white dwarf-main sequence PCEB IK
Peg \citep{Davis2010}. This indicates that an energy source in
addition to gravitational energy is being exploited during the CE
ejection process that is not accounted for in the standard energy
budget formulation such as thermal energy and recombination energy of
ionized material within the giant's envelope
\citep{Han2002,Dewi2000,Webbink2008}.

The second issue is that previous studies into the formation and evolution
of white dwarf binaries \citep[e.g.][]{deKool1993, Willems2004} have
assumed that $\alpha_{\rmn{CE}}$ is a global constant,
i.e. $\alpha_{\rmn{CE}}$ is the same for all progenitor systems,
irrespective of their parameters upon entering the CE phase. This may be a
somewhat na\"{i}ve assumption. Indeed, \citet{Terman1996} suggest that
$\alpha_{\rmn{CE}}$ may depend upon the internal structure of the Roche
lobe filling primary progenitor.

In this spirit, \citet{Politano2007} calculated the present-day
population of PCEBs and CVs if $\alpha_{\rmn{CE}}$ is a function of
the secondary mass. They considered both a power-law dependence
(i.e. $\alpha_{\rmn{CE}}\propto{M_{2}}^{n}$, where $n$ is some power),
and a dependence of $\alpha_{\rmn{CE}}$ on a cut-off mass. If the
secondary mass was below this cut-off limit, then a merger between the
two stellar components is assumed to be unavoidable (formally
$\alpha_{\rmn{CE}}=0$ in this case).

As \citet{Davis2010} found, however, these formalisms tend to
underestimate $\alpha_{\rmn{CE}}$ for PCEBs with low-mass secondaries
($\la{0.2}$ M$_{\odot}$). For example even though a $n=2$ dependence
could account for IK Peg, it predicts that systems with
$M_{2}\la{0.4}$ M$_{\odot}$ cannot survive the CE phase. This is in
conflict with observations that show that PCEBs with secondary masses
as low as $M_{2}\approx{0.05}$ M$_{\odot}$ do exist.

During the preparation of this paper, we became aware of similar studies
being carried out by \citet{Zorotovic2010} and
\citet{DeMarco2011}. \citet{Zorotovic2010} applied the CE reconstruction
method described by \citet{Nelemans2005} to the observed sample of
PCEBs. They investigated whether $\alpha_{\rmn{CE}}$ depended on either the
secondary mass or the orbital period immediately after the CE phase. They
concluded that there was no dependence in either case. However, this
conclusion was based on a qualitative, `by-eye', analysis. 

\citet{DeMarco2011}, on the other hand, found evidence that
$\alpha_{\rmn{CE}}$ increases with increasing mass of the progenitor
primary star, but decreases with increasing mass of the secondary. However,
they only considered PCEBs which underwent negligible angular momentum loss
since emerging from the CE phase, which limited their sample size to 30
systems.

Our investigation is similar to these studies, except that we investigate
whether $\alpha_{\rmn{CE}}$ can be treated as a first-order function of a
combination of the binary parameters -- white dwarf mass, progenitor
primary mass and orbital period at the start of the CE phase, secondary
mass and post-CE orbital period -- by performing multi-regression analysis.
Our method for reconstructing the CE phase for the observed PCEB sample is
similar to that used by \citet{Zorotovic2010}, but has been independently
developed. The aim is to determine whether correlations exist between
$\alpha_{\rmn{CE}}$ and the binary parameters. The resulting fits will then
be useful recipes for binary evolution and population synthesis codes. Such
fits may also be useful to CE theoreticians in order to inform and
constrain their models.

Our PCEB sample consists of systems contained in the \citet{Ritter2003}
catalogue, Edition 7.14 (2010) and the 34 new PCEBs detected from the Sloan
Digital Sky Survey (SDSS), in contrast to \citet{DeMarco2011}. For an
in-depth review of this sample, we refer the reader to
\citet{Zorotovic2010}.

The structure of the paper is as follows. In Section 2 we describe our
method for reconstructing the CE phase for the observed sample of
PCEBs. Our results are then given in Section 3, which are then discussed in
Section 4. We then compare our work with similar studies carried out by
\citet{Zorotovic2010} and \citet{DeMarco2011} in Section 5. Finally, our
conclusions are given in Section 6.

\section{Computational Method}
\label{section:method}

In the following subsections, we describe the energy budget equation
describing the CE phase in more detail. We then explain how we can
calculate the orbital separation of the system immediately after the
CE phase from the cooling age of the white dwarf and the assumed
angular momentum loss rate. Finally, we discuss how we use our
population synthesis code BiSEPS, to evolve the possible progenitors
of a given observed system through the CE phase. We then use the
energy budget equation (c.f. eqn. \ref{Ebind_1}) to solve for
$\alpha_{\rmn{CE}}$.

\subsection{Calculating $\alpha_{\rmn{CE}}$}
\label{subsection:alpha}

We calculate the binding energy of the primary's envelope,
$E_{\rmn{bind}}$, using:
\begin{equation}
E_{\rmn{bind}}=-\frac{GM_{1}M_{\rmn{env}}}{\lambda R_{1}},
\label{Ebind_4}
\end{equation}
where $M_{1}$ is the mass of the progenitor primary, $M_{\rmn{env}}$
is the envelope mass and $R_{1}$ is the radius of the primary, which
is also equal to its Roche lobe radius at the start of the CE
phase. The parameter $\lambda$ is the ratio between the approximate
expression of the gravitational binding energy given by
eqn. (\ref{Ebind_4}) and the exact expression:
\begin{equation}
E_{\rmn{bind}}=-\int^{M_{1}}_{M_{\rmn{c}}}\frac{GM_{1}(r)}{r}\rmn{d}m,
\label{Ebind_g}
\end{equation}
where $M_{\rmn{c}}$ is the core mass (i.e. the mass of the proto-white
dwarf) and $M_{1}(r)$ is the mass enclosed within a radius
$r$. Alternatively, if we consider the internal energy of the
envelope, we have
\begin{equation}
E_{\rmn{bind}}=-\int^{M_{1}}_{M_{\rmn{c}}}\frac{GM_{1}(r)}{r}\rmn{d}m+\alpha_{\rmn{th}}\int^{M_{1}}_{M_{\rmn{c}}}U\rmn{d}m,
\label{Ebind_b}
\end{equation}
where $U$ is the energy per unit mass, and $\alpha_{\rmn{th}}$ is the
fraction of thermal energy which is used to unbind the envelope
\citep{Dewi2000}. We calculate $\alpha_{\rmn{CE}}$ for observed systems by
adopting either eqn. (\ref{Ebind_g}) ($\lambda=\lambda_{\rmn{g}}$) or
eq. (\ref{Ebind_b}) ($\lambda=\lambda_{\rmn{b}}$) to calculate $\lambda$ in
eqn. (\ref{Ebind_4}). Furthermore, we follow \citet{Dewi2000} and assume
that $\alpha_{\rmn{th}}=1$ for this case.

For a binary system which has an initial orbital separation
$A_{\rmn{CE,i}}$ immediately before the CE phase (i.e. at the point
when the primary giant just fills its Roche lobe), and an orbital
separation $A_{\rmn{CE,f}}$ immediately after the CE phase (i.e. the
point when the PCEB emerges from the CE phase), the change in orbital
energy is given by:
\begin{equation}
\Delta
E_{\rmn{orb}}=\frac{GM_{\rmn{c}}M_{2}}{2A_{\rmn{CE,f}}}-\frac{GM_{1}M_{2}}{2A_{\rmn{CE,i}}},
\label{Eorb_5}
\end{equation}
where $M_{2}$ is the mass of the secondary star.

By combining eqns. (\ref{Ebind_1}), (\ref{Ebind_4}) and (\ref{Eorb_5})
we can solve for $\alpha_{\rmn{CE}}$. This gives,
\begin{equation}
\alpha_{\rmn{CE}}=\frac{2M_{\rmn{env}}M_{1}}{r_{1}M_{2}\lambda}\left(\frac{A_{\rmn{CE,f}}}{M_{\rmn{c}}A_{\rmn{CE,i}}-M_{1}A_{\rmn{CE,f}}}\right),
\label{alpha_6}
\end{equation}
where $r_{1}=R_{1}/A_{\rmn{CE,i}}$ is the radius of the primary star
in units of the initial orbital separation.

\subsection{Calculating $A_{\rmn{CE,f}}$ for observed systems}
\label{subsection:postCEsep}

Generally, the observed orbital separation of a PCEB is not the same
as the orbital separation immediately after the CE phase. This is
because the PCEB would have undergone orbital period evolution due to
angular momentum losses via gravitational radiation or magnetic
braking since emerging from the CE phase. We therefore follow
\citet{Schreiber2003} to calculate the orbital period of our observed
sample of PCEBs immediately after their emergence from the CE phase,
$P_{\rmn{CE}}$, using the cooling time of the white dwarf,
$t_{\rmn{cool}}$.

We apply the disrupted magnetic braking paradigm \citep{Spruit1983,
  Rappaport1983} and assume that magnetic braking is ineffective for
secondaries that have masses less than the fully convective mass limit
for main sequence stars, $M_{\rmn{conv,MS}}=0.35$ M$_{\odot}$
\citep{Hurley2002}. Thus, for observed PCEBs with
$M_{2}\le{M_{\rmn{conv,MS}}}$, we assume that the systems are driven
purely by gravitational wave radiation. We therefore calculate
$P_{\rmn{CE}}$ according to equation (8) of \citet{Schreiber2003}:
\begin{eqnarray} 
P_{\rmn{CE}}^{8/3}=\frac{256G^{2/3}(2\pi)^{8/3}t_{\rmn{cool}}}{5c^5}\nonumber\\
\times \frac{M_{1}M_{2}}{(M_{1}+M_{2})^{1/3}}+P_{\rmn{orb}}^{8/3}.
\label{tcool_gr}
\end{eqnarray}

The evolution of observed systems with $M_{2}>M_{\rmn{conv,MS}}$ will
be driven by a combination of gravitational radiation and magnetic
braking. In this investigation, we adopt the magnetic braking
formalism given by \citet{Hurley2002}. However, we can neglect the
contribution due to gravitational radiation because the associated
angular momentum loss rate is much less than that associated with
magnetic braking. To calculate $P_{\rmn{CE}}$ we therefore apply
\begin{eqnarray}
P_{\rmn{CE}}^{10/3}=\frac{10(2\pi)^{10/3}\eta_{\rmn{h}}t_{\rmn{cool}}R_{2}^{3}M_{\rmn{conv}}}{1.72\times{10}^{15}G^{2/3}}\nonumber\\
 \times\frac{(M_{1}+M_{2})^{1/3}}{M_{1}M_{2}^{2}}+P_{\rmn{orb}}^{10/3}
\label{tcool_hurley}
\end{eqnarray}
where $P_{\rmn{CE}}$, $P_{\rmn{orb}}$ and $t_{\rmn{cool}}$ are
expressed in yr, and $G$ is expressed in R$_{\odot}^{3}$
M$_{\odot}^{-1}$ yr$^{-2}$. We also include a normalisation factor
$\eta_{\rmn{h}}=0.17$ in order to obtain an angular momentum loss rate
that is appropriate for the observed width and location of the CV
period gap \citep{Davis2008}. We estimate the convective envelope mass
of the secondary star as
\begin{equation}
  \frac{M_{\rmn{conv}}}{\rmn{M}_{\odot}} = \left\{
  \begin{array}{l l}

  0 & \quad \mbox{$M_{2}>{1.25}$ M$_{\odot}$}\\
  0.35\left(\frac{1.25-M_{2}}{0.9}\right)^{2} & \quad \mbox{$0.35\le{M_{2}/\rmn\
{M}_{\odot}}\le{1.25}$}\\

  \end{array} \right.
\label{Mconv}
\end{equation}
\citep{Hurley2002}. Note that equation (\ref{Mconv}) calculates the
convective envelope mass for a star on the zero-age main sequence
(ZAMS). In reality, the mass of the convective envelope is a function of
the fraction of the main sequence lifetime that the star has passed through
\citep[e.g.][]{Hurley2000}, thus making equation (\ref{Mconv}) time
dependent and the solution to $P_{\rmn{CE}}$ in equation
(\ref{tcool_hurley}) non-trivial.

\subsection{The modified BiSEPS code}
\label{subsection:BiSEPS}

We use a modified version of the BiSEPS code
\citep{Willems2002,Willems2004} to evolve all possible progenitors of each
observed PCEB through the CE phase, and to calculate $\alpha_{\rmn{CE}}$
for each of them.

The BiSEPS code employs the single star evolution formulae by
\citet{Hurley2000} and a binary evolution scheme based on that described by
\citet{Hurley2002}. We evolve a large number of binary systems initially
consisting of two ZAMS stellar components. The stars are assumed to have a
Population I chemical composition and the orbits are circular at all
times. The initial secondary mass is equal to the observed secondary mass
of the observed PCEB we are considering. The initial primary masses and
orbital periods are obtained from a two-dimensional grid consisting of 1000
logarithmically spaced points in both primary mass and orbital period. The
initial primary masses are in the range of 0.1 to 20 M$_{\odot}$, while the
initial orbital periods are in the range of 0.1 to 100\,000 d. Hence, we
evolve $1\times{10}^{6}$ binaries for a maximum evolution time of 10
Gyr. For symmetry reasons only systems where $M_{2}>M_{1}$ are evolved.

Prior to the onset of the CE phase, the primary star will lose mass
via stellar winds. For stars on and after the main sequence, mass loss
rates are calculated using the formalism described by
\citet{Kudritzki1978}. The super-wind phase on the AGB is calculated
using the prescription described by \citet{Vassiliadis1993}.

If a binary configuration undergoes a CE phase, and the mass of the
primary core is equal to the observed white dwarf mass within the
observed uncertainty \footnote{We can assume that the core of the
  primary giant is equal to the white dwarf mass because the increase
  in the core's mass during the CE phase will be negligible.} then we
can use eqn. (\ref{alpha_6}) to calculate $\alpha_{\rmn{CE}}$ for that
system. For each progenitor, we calculate $\lambda_{\rmn{g}}$ or
$\lambda_{\rmn{b}}$ by linear interpolating between values tabulated
by \citet{Dewi2000}. We extended this table using full stellar models
calculated by the Eggleton code (provided by Marc van der Sluys,
private communication) and the EZ stellar evolution code
\citep{Paxton2004}. We refer the reader to \citet{Davis2010} for
further details.

For a given observed PCEB, we now have a range of possible values of
$\alpha_{\rmn{CE}}$, each corresponding to a possible progenitor system. A
value of $\alpha_{\rmn{CE}}$ is given a weighting equal to the formation
probability of the associated progenitor system. The formation probability
for a progenitor can be found from the initial mass function (IMF),
$f(M_{1,\rmn{i}})$, where $M_{1,\rmn{i}}$ is the ZAMS primary mass (see
eqn. \ref{IMF}), and from the initial orbital period distribution
$H(P_{\rmn{orb,i}})$, which can be found by using the initial orbital
separation distribution (IOSD), $h(a_{\rmn{i}})$ (see eqn. \ref{IOSD}), via
Kepler's Third Law. The quantities $P_{\rmn{orb,i}}$ and $a_{\rmn{i}}$ are
the initial orbital period and orbital separation at the ZAMS stage
respectively.

If the internal energy of the envelope is considered
(i.e. $\lambda=\lambda_{\rmn{b}}$), \citet{Dewi2000} find that the binding
energy of the envelope is positive if the radius of the primary giant star
is sufficiently large, formally indicated by a value of
$\lambda_{\rmn{b}}<0$. This is a result of the fact that the magnitude of
the radiation pressure exceeds the gravitational force. As a result, the
prescription given by eqn. (\ref{alpha_6}) breaks down, and we cannot
follow the evolution of such progenitors through the CE phase using the
energy budget formalism (the implications of $\lambda_{\rmn{b}}<0$ is
discussed in more detail in Section \ref{section:discussion}). Therefore,
progenitor binaries of a given PCEB with such primary giants are discounted
in the aforementioned weighting calculation.

For the secondary mass, we just consider a flat initial mass ratio
distribution, $g(q_{\rmn{i}})=1$ (see eqn. \ref{IMRD}), where
$q_{\rmn{i}}=M_{2,\rmn{i}}/M_{1,\rmn{i}}$ \footnote{Note that it does
  not matter how we determine the formation probability of the
  secondary star, as it will be the same for all progenitors of a
  given PCEB.}, where $M_{2,\rmn{i}}$ is the mass of the ZAMS
secondary. Hence, the probability that a PCEB formed from a CE phase
with an ejection efficiency in the range $\Delta \alpha_{\rmn{CE}}$ is
proportional to $f(M_{1,\rmn{i}})h(a_{\rmn{i}})g(q_{\rmn{i}})$.

\begin{figure}
  \includegraphics[scale=0.43]{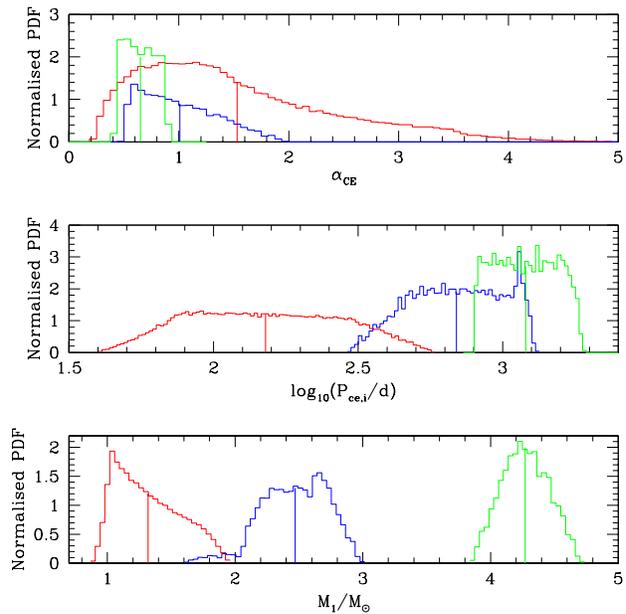}
  \caption{Distributions in the possible primary progenitor mass at the
    start of the CE phase ($M_{1}$, bottom panel), orbital period at the
    start of the CE phase, $P_{\rmn{CE,i}}$, (middle panel) and the CE
    ejection efficiency, $\alpha_{\rmn{CE}}$ (where we use
    $\lambda=\lambda_{\rmn{g}}$; top panel) for SDSS2216+0102 (red),
    SDSS1548+4057 (blue) and for SDSS0303-0054 (green). The vertical lines
    indicate the mean values for each histogram.}
  \label{examples}
\end{figure}

\begin{figure*}
  \begin{minipage}{175mm}
    \begin{center}
      \includegraphics[scale=0.80]{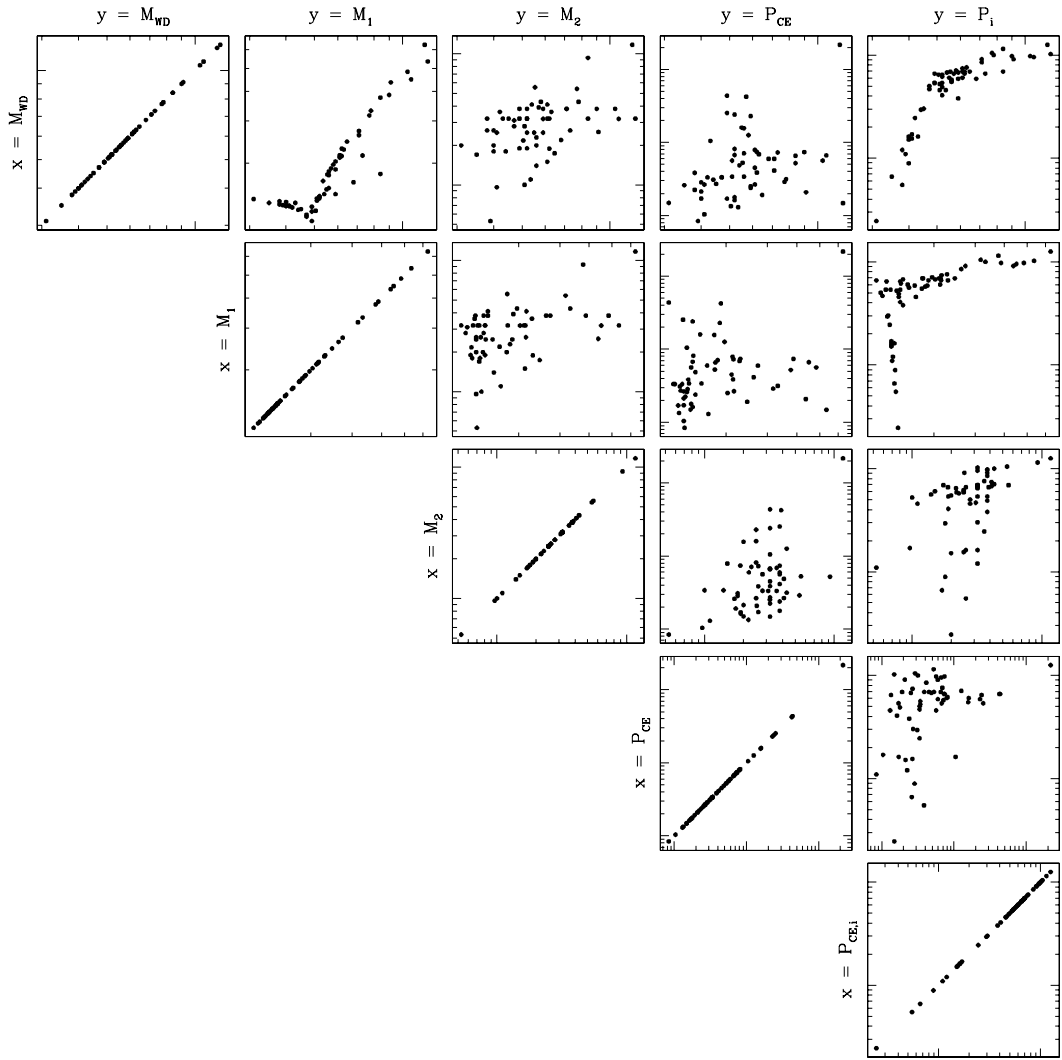}
      \caption{Correlation matrix plot for the considered binary
        parameters. Rows correspond to independent variables, while columns
        correspond to dependent variables. Errors and numbers are omitted
        for clarity.}
      \label{corr_matrix}
   \end{center}
  \end{minipage}
\end{figure*}

Fig. \ref{examples} shows distributions of possible primary progenitor mass
at the start of the CE phase ($M_{1}$, bottom panel), orbital period at the
onset of the CE phase ($P_{\rmn{CE,i}}$, middle panel) and the CE ejection
efficiency ($\alpha_{\rmn{CE}}$, top panel, where we use
$\lambda=\lambda_{\rmn{g}}$) for SDSS2216+0102 (red), SDSS1548+4057 (blue)
and SDSS0303-0054 (green). The vertical lines indicate the mean value for
each histogram.

The white dwarf masses for SDSS2216+0102, SDSS1548+4057 and
SDSS0303-0054 are $0.400\pm{0.060}$ M$_{\odot}$, $0.646\pm{0.032}$
M$_{\odot}$ and $0.912\pm{0.034}$ M$_{\odot}$ respectively. We can see
from the bottom and middle panels that the average primary mass and
orbital period immediately before the CE phase increase for increasing
mass of the white dwarf descendant.

The orbital periods of these systems immediately after the CE phase
are approximately the same ($\approx{0.2}$ d). Therefore, the typical
progenitor of SDSS0303-0054 would need to spiral-in further during the
CE phase than SDSS2216+0102 in order to reach the same post-CE orbital
period. Therefore, for the former system, more orbital energy would be
needed to eject the CE from the system than for the latter
system. Hence, the typical ejection efficiency will be less efficient
for SDSS0303-0054 than for SDSS2216+0102. This is illustrated by the
histograms in the top panel. The average orbital separation and
primary mass before the CE phase for SDSS1548+4057 lies between the
average values for SDSS2216+0102 and SDSS0303-0054. Therefore, the
average ejection efficiency for this system also lies between
SDSS2216+0102 and SDSS0303-0054.

\begin{figure*}
  \begin{minipage}{175mm}
    \begin{center}
      \includegraphics[scale=0.80]{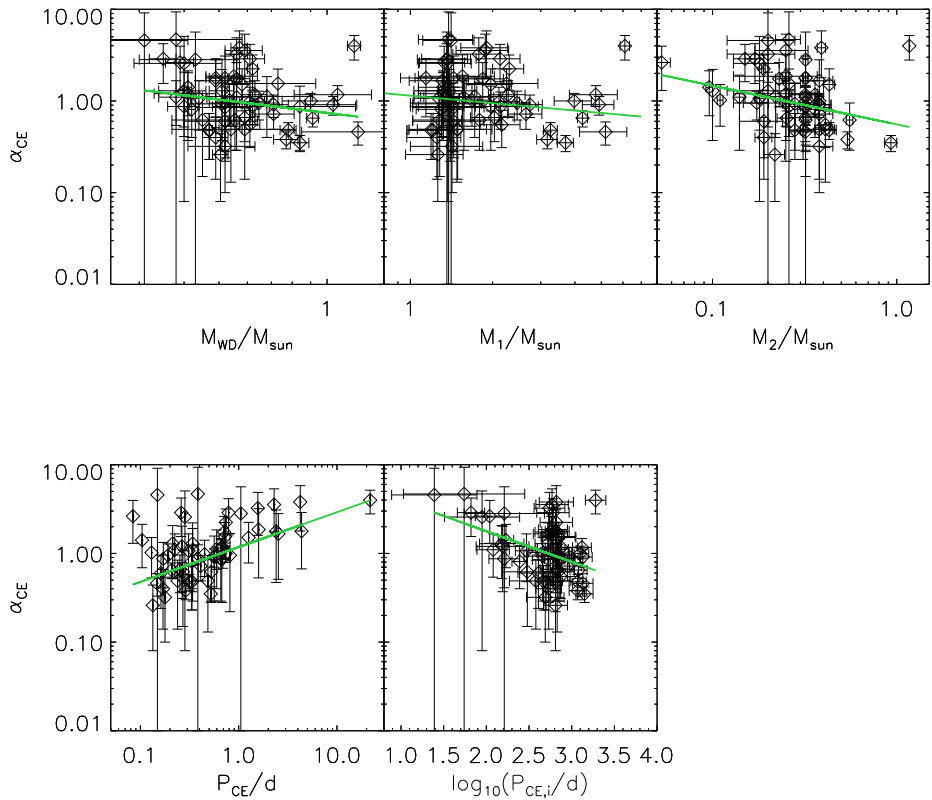}
      \caption{Our reconstructed values of $\alpha_{\rmn{CE}}$ versus the
        following binary parameters, where we take
        $\lambda=\lambda_{\rmn{g}}$: white dwarf mass, $M_{\rmn{WD}}$;
        primary mass at the moment the CE phase commences, $M_{1}$; secondary
        mass, $M_{2}$; orbital period at the moment the CE ends,
        $P_\rmn{CE}$; log$_{10}$ of the orbital period at the start of the CE
        phase, $P_{\rmn{CE,i}}$. The solid green lines indicate linear fits to
        the data.}
      \label{ace_params_G}
  \end{center}
  \end{minipage}
\end{figure*}

\begin{figure}
  \begin{center}
    \includegraphics[scale=0.5]{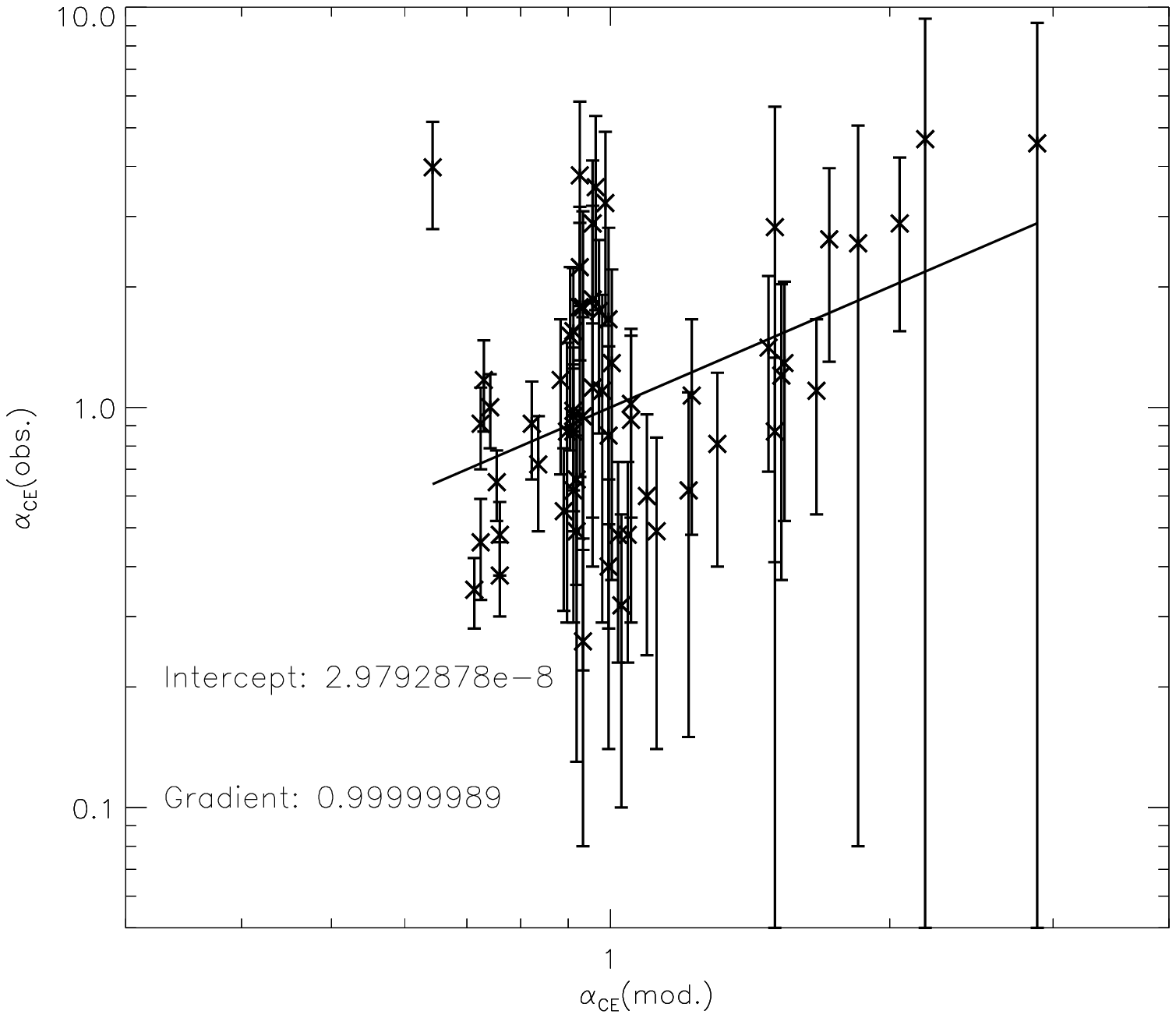}
    \caption{Our reconstructed values of $\alpha_{\rmn{CE}}$ (y-axis,
      $\alpha_{\rmn{CE}}(\rmn{obs.})$) versus values of
      $\alpha_{\rmn{CE}}$ (x-axis, $\alpha_{\rm{CE}}(\rmn{mod.})$) which
      have been calculated from eqn. (\ref{progenitor_model_G}). Also shown
      is the value of the intercept and gradient of the linear fit through
      the data, which is shown as the solid black line.}
    \label{fig_progenitor_model_G}
    \includegraphics[scale=0.5]{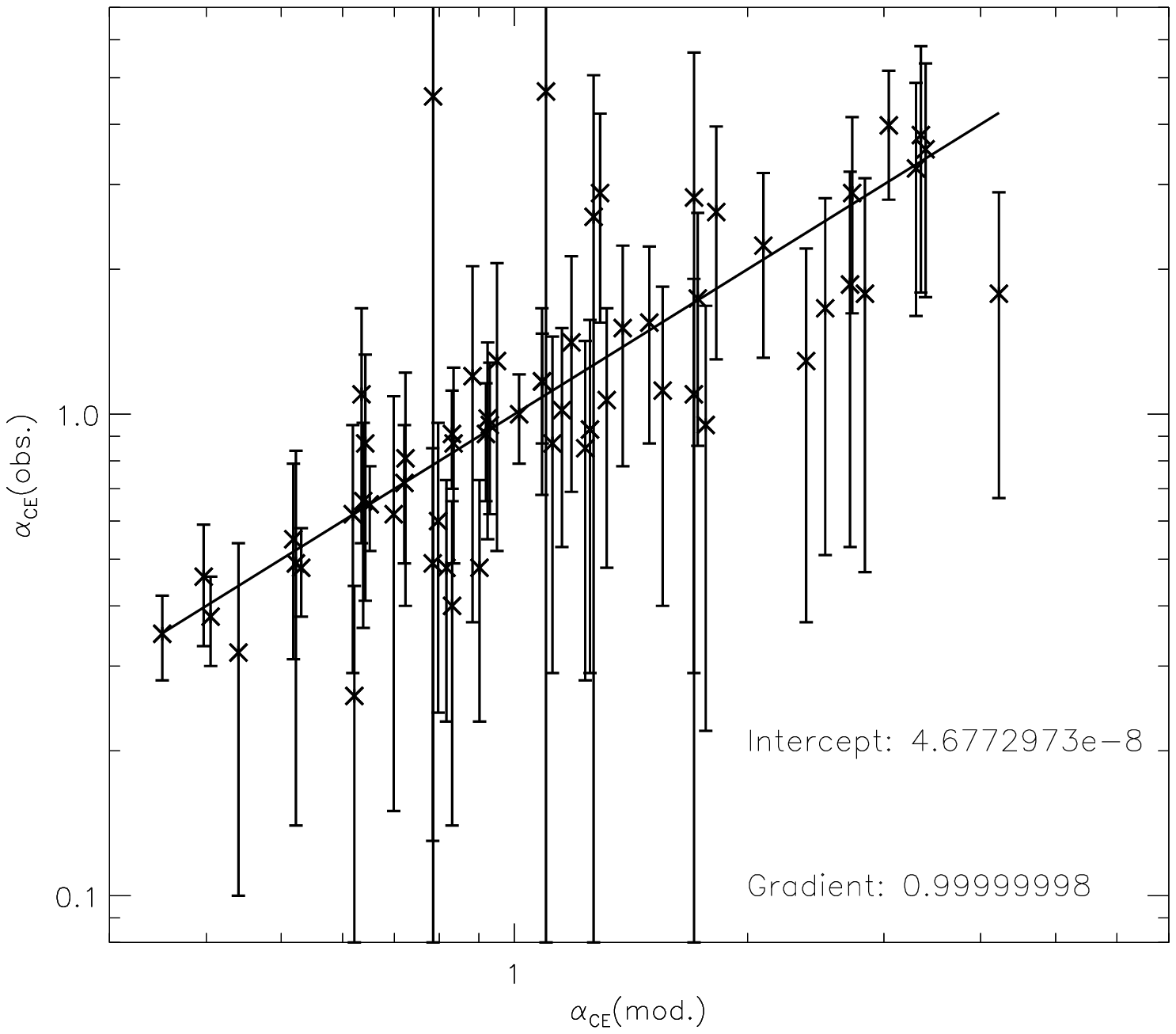}
    \caption{Same as Fig. \ref{fig_progenitor_model_G}, except
      $\alpha_{\rmn{CE}}(\rmn{mod.})$ has been calculated from
      eqn. (\ref{PCEB_model_G}).}
    \label{fig_PCEB_model_G}
  \end{center}
\end{figure}

\begin{table*}
  \begin{center}
  \begin{minipage}{95mm}
    \caption{A summary of the fit parameters $\epsilon_{0}$ and
      $\epsilon_{1}$ for the data in each panel of Fig. \ref{ace_params_G},
      where we consider $\lambda=\lambda_{\rmn{g}}$. Also shown for each
      panel are the values of $\mathcal{F}$ and the probability of
      exceeding $\mathcal{F}$, $\mathscr{P}(\mathcal{F})$. Finally, an
      indication of whether we can reject the null hypothesis
      $\mathscr{H}^{\mathcal{F}}_{0}$ is included (see text for further
      details).}
    \begin{tabular}{@{}lccccc@{}}
      \hline
      $Q$ & $\epsilon_{0}$  & $\epsilon_{1}$  &
      $\mathcal{F}$  & $\mathscr{P}(\mathcal{F})$ &  Reject
      $\mathscr{H}^{\mathcal{F}}_{0}$ ?$\,^{\rmn{a}}$  \\
      \hline
      $M_{\rmn{WD}}$ & -0.13$\pm$0.05  & -0.48$\pm$0.27  &  3.2  & $1\times{10}^{-1}$ &  No \\ 
      $M_{1}$ & 0.06$\pm$0.09  &  -0.27$\pm$0.18  &  2.2  & $1\times{10}^{-1}$ &  No \\ 
      $M_{2}$ & -0.26$\pm$0.08 &  -0.42$\pm$0.15  &  7.4 & $9\times{10}^{-3}$  &  Yes \\
      $P_{\rmn{CE}}$ & 0.07$\pm$0.04  &  0.39$\pm$0.07  &  29.5  & $<10^{-6}$ &  Yes \\
      $P_{\rmn{CE,i}}$ & 0.94$\pm$0.33  &  -0.34$\pm$0.11 &  9.1 & 4$\times{10}^{-3}$ &  Yes \\
      \hline
    \end{tabular}
    $^{\rmn{a}}$ $\mathcal{F}_{\rmn{crit}}=4.00$ for a 0.05 significance level. \\
    \label{fits_g}
  \end{minipage}
  \end{center}
\end{table*}

\begin{table*}
  \begin{center}
  \begin{minipage}{155mm}
    \caption{A summary of our fit parameters, obtained from a
      multi-regression analysis of the values of $\alpha_{\rmn{CE}}$ (where
      $\lambda=\lambda_{\rmn{g}}$) as a function of two or more binary
      parameters. Shown are the coefficients $\beta_{0...m}$, the
      F-statistic, $\mathcal{F}$, and the probability of exceeding
      $\mathcal{F}$, $\mathscr{P}(\mathcal{F})$. We also indicate whether
      we can reject the null hypothesis $\mathscr{H}_{0}^{\mathcal{F}}$.}
    \begin{tabular}{@{}lcccccccc@{}}
      \hline
      Variables &  $\beta_{0}$  &  $\beta_{1}$  &  $\beta_{2}$  &
      $\beta_{3}$  &  $\beta_{4}$  &  $\mathcal{F}$  &
      $\mathscr{P}(\mathcal{F})$ &  Reject $\mathscr{H}_{0}^{\mathscr{F}}$ ? \\  
      \hline
      \multicolumn{9}{c}{Progenitor parameters} \\
      $P_{\rmn{CE,i}}+M_{\rmn{WD}}$ &  2.35 &  -0.78$\pm$0.25 & 1.09$\pm$0.56
      & - & - & 3.6 & 6$\times{10}^{-2}$  &  No\\
      $+M_{2}$ & 1.94 & -0.67$\pm$0.27 &
      1.04$\pm$0.56 & -0.19$\pm$0.19 & - & 1.1 & $3\times{10}^{-1}$  &
      No  \\  
      $+M_{1}$ & 1.88 &
      -0.66$\pm$0.30 &  0.97$\pm$1.38 & -0.20$\pm$0.20 & 0.04$\pm$0.72 &
      $0.003$ & $9.6\times{10}^{-1}$  &  No\\
      \hline
      \multicolumn{9}{c}{Observable (PCEB) parameters} \\
      $P_{\rmn{CE}}+M_{\rmn{WD}}$ & -0.03 & 0.20$\pm$0.03 &
      -0.85$\pm$0.21 & - & - & 16.6 & $2\times{10}^{-4}$  &  Yes \\
      $+M_{2}$ & -0.31 & 0.28$\pm$0.02 &
      -0.95$\pm$0.13 & -0.17$\pm$0.09 & - & 38.4 & $<1\times{10}^{-6}$
      &  Yes \\ 
      
      \hline
    \end{tabular}
    \label{regress_g}
  \end{minipage}
  \end{center}
\end{table*}

\subsection{Statistical Approach}
\label{subsection:stats}

Our aim in this investigation is to apply a chi-squared multi-regression
analysis (via the Levenberg-Marquardt method) to obtain fits of
$\alpha_{\rmn{CE}}$ for two cases; we do this by first determining whether
correlations exist between $\alpha_{\rmn{CE}}$ and the binary progenitor
parameters at the point that the CE phase commences (i.e. the moment the
primary progenitor star just fills its Roche lobe), which will be useful
for population synthesis calculations. In the second case, we determine
whether correlations exist between $\alpha_{\rmn{CE}}$ and the (observable)
PCEB parameters, which may be a useful diagnostic to probe CE physics.

We proceed as follows. We fit our reconstructed values of
$\alpha_{\rmn{CE}}$ according to the equation
\begin{equation}
\rmn{log}_{10}\alpha_{\rmn{CE}}=\beta_{0}+\sum_{i=1}^{m}\beta_{i}\rmn{log}_{10}Q_{i},
\label{multi_fit}
\end{equation}
where $\beta_{0...m}$ are constants. If we are considering the progenitor
binary parameters only, then
$Q\in{\{P_{\rmn{CE,i}},M_{2},M_{\rmn{WD}},M_{1}\}}$, with $m=4$. If we
consider the PCEB parameters, then
$Q\in{\{P_{\rmn{CE}},M_{2},M_{\rmn{WD}}\}}$, with $m=3$.

Strictly speaking, we should only apply the Levenberg-Marquardt chi-squared
method if the errors are normally distributed. As we can see from
Fig. \ref{examples}, this is not the case. We may therefore under-estimate
the uncertainties in our fit parameters. To tackle this, we rescale the
standard deviations for the values of $\alpha_{\rmn{CE}}$ such that we
obtain a reduced chi-squared value of 1.

To determine which variable to include first in our fit, we perform a
1-dimensional (1D) linear fit of our reconstructed values of
$\alpha_{\rmn{CE}}$ versus each of the binary parameters. Specifically, we
fit the function of the form
\begin{equation}
\rmn{log}_{10}\alpha_{\rmn{CE}}=\epsilon_{0}+\epsilon_{1}\rmn{log}Q,
\label{linr}
\end{equation}
where $\epsilon_{0}$ and $\epsilon_{1}$ are constants.

Then, to determine if a linear fit -- as opposed to a constant fit -- to
the data is actually warranted (i.e. we reject the null hypothesis
$\mathscr{H}_{0}^{\mathcal{F}}$), we perform the F-test (see Section
\ref{subsection:F_test}). We choose a significance level of
$\alpha=0.05$. The first variable which we include in our fit is the one
which gives the smallest probability, $\mathscr{P}(\mathcal{F})$, that we
will obtain an F-statistic smaller than the obtained value, $\mathcal{F}$.

Once we have added our first variable to the fit, we determine the next
binary parameter to include, by performing both a constant and a 1D fit to
the residuals as a function of the remaining binary parameters. The next
variable which we add is the one which gives the smallest value of
$\mathscr{P}(\mathcal{F})$. In principle, we can repeat this process until
we have a fit in terms of all binary parameters.

However, special care should be taken when performing regression-analysis;
strictly speaking, the predictor variables should be independent of one
another. We check whether there is any correlation between the binary
parameters via a correlation matrix plot as shown in
Fig. \ref{corr_matrix}. There are clearly strong correlations between
$M_{\rmn{WD}}$, $M_{1}$ and $P_{\rmn{CE,i}}$. This is to be expected
because the core mass of a Roche lobe-filling star is a (weak) function of
the total mass, and as a function of the orbital period via the star's
Roche lobe radius and Kepler's Third Law. Hence, by including some or all
of these variables we may over-fit the data, and we may render certain
variables redundant in the fit.

As we add each variable to our fit as described above, we once again apply
the $\mathcal{F}$-Test to determine whether the fit is a statistically
better fit than the one without this added binary parameter.

Our results for the case with $\lambda=\lambda_{\rmn{g}}$ and
$\lambda=\lambda_{\rmn{b}}$ are discussed in turn in the following
Sections.

\section{Results}
\label{section:results}

The observed binary parameters (component masses and orbital period) as
well as the cooling age of the white dwarf, $t_{\rmn{cool}}$, are
summarised in table \ref{CERecon_SDSS} for PCEBs detected by the
SDSS. Table \ref{CERecon_RKCat} is the same but for PCEBs recorded in the
\citet{Ritter2003} catalogue, Edition 7.14 (2010). The average values of
$\alpha_{\rmn{CE}}$, $P_{\rmn{CE,i}}$ and $M_{1}$ and their standard
deviations (calculated from distributions similar to those illustrated in
Fig. \ref{examples}) for each observed PCEB are also summarised in Table
\ref{CERecon_SDSS} and \ref{CERecon_RKCat} where we do not consider the
internal energy of the progenitor primary envelope. Tables
\ref{CERecon_SDSS_B} and \ref{CERecon_RKCat_B} give the average values of
$M_{1}$, $P_{\rmn{CE,i}}$ and $\alpha_{\rmn{CE}}$ for each observed PCEB
when we do consider the internal energy of the progenitor primary envelope.

\subsection{For $\lambda=\lambda_{\rmn{g}}$}

\begin{table*}
  \begin{center}
  \begin{minipage}{95mm}
    \caption{Same as Table \ref{fits_g} except for the data shown in
      Fig. \ref{ace_params_B}.}
    \begin{tabular}{@{}lccccc@{}}
      \hline
      $Q$ & $\epsilon_{0}$  & $\epsilon_{1}$  &
      $\mathcal{F}$ & $\mathscr{P}(\mathcal{F})$ &  Reject
      $\mathscr{H}^{\mathcal{F}}_{0}$ ?$^a$  \\
      \hline
      $M_{\rmn{WD}}$ & -1.03$\pm$0.10  &  -2.10$\pm$0.34  &  36.1 & $<10^{-6}$ &  Yes \\
      $M_{1}$ & -0.21$\pm$0.07  &  -1.26$\pm$0.23  &  30.0 & $<10^{-6}$  &  Yes \\
      $M_{2}$ & -0.89$\pm$0.13  &  -0.70$\pm$0.20  &  2.9 & $6\times{10}^{-4}$  &  Yes \\
      $P_{\rmn{CE}}$ & -0.45$\pm$0.06  &  0.15$\pm$0.10   &  2.7 & $1\times{10}^{-1}$  &  No \\
      $P_{\rmn{CE,i}}$ & 1.55$\pm$0.43  &  -0.76$\pm$0.16   & 22.4 &  $ 2\times{10}^{-5}$  &   Yes  \\  
      \hline
    \end{tabular}

    $^{\rmn{a}}$ $\mathcal{F}_{\rmn{crit}}=4.00$ for a 0.05 significance level. \\

    \label{fits_b}
  \end{minipage}
  \end{center}
\end{table*}

\begin{table*}
  \begin{center}
  \begin{minipage}{155mm}
    \caption{The same as Table \ref{regress_g} except now for our data
      calculated using $\lambda_{\rmn{b}}$.}
    \begin{tabular}{@{}lcccccccc@{}}
      \hline
      Variables  &  $\beta_{0}$  &  $\beta_{1}$  &  $\beta_{2}$  &
      $\beta_{3}$  &  $\beta_{4}$  &  $\mathcal{F}$  &
      $\mathscr{P}(\mathcal{F})$ &  Reject $\mathscr{H}_{0}^{\mathscr{F}}$ ?\\  
      \hline
      \multicolumn{9}{c}{Progenitor parameters} \\
      $M_{\rmn{WD}}+M_{2}$ & -1.08 & -1.94$\pm$0.29 &
      -0.14$\pm$0.14 & - & - & 0.4 & 0.5 &  No \\
      $+P_{\rmn{CE,i}}$ & -0.54 & -1.65$\pm$0.39 &
      -0.13$\pm$0.15 & -0.17$\pm$0.16 & - & 0.6 & 0.4 & No \\  
      $+M_{1}$ & 0.98 &
      0.22$\pm$1.26 &  -0.07$\pm$0.15 & -0.46$\pm$0.24 & -0.98$\pm$0.63 &
      1.1 & 0.3 & No \\
      \hline
      \multicolumn{9}{c}{Observable (PCEB) parameters} \\
      $M_{\rmn{WD}}+P_{\rmn{CE}}$ & -1.18 & -3.20$\pm$0.26 & 0.22$\pm$0.03 & - &
      - & 71.8 & $<{10}^{-6}$ & Yes  \\
      $+M_{2}$ & -1.53 & -2.50$\pm$0.16 &
      0.31$\pm$0.02 & -1.02$\pm$0.09 & - & 127.5 & $<{10}^{-6}$ & Yes \\

      \hline
    \end{tabular}
    \label{regress_b}
  \end{minipage}
  \end{center}
\end{table*}

Our values of $\alpha_{\rmn{CE}}$ versus $M_{\rmn{WD}}$, $M_{1}$, $M_{2}$,
$P_{\rmn{CE}}$ and $P_{\rmn{CE,i}}$ are shown in Fig. \ref{ace_params_G}
for the case where we consider $\lambda=\lambda_{\rmn{g}}$. Each point
corresponds to the mean value of $\alpha_{\rmn{CE}}$, while the vertical
lines correspond to the standard deviation.

The values of $\epsilon_{0}$, $\epsilon_{1}$ for each panel of
Fig. \ref{ace_params_G} are given in Table \ref{fits_g}. The solid green
curves in Fig. \ref{ace_params_G} are the linear fits. For comparison, we
also fit a constant function to the data in Figs. \ref{ace_params_G} of the
form
\begin{equation}
\rmn{log}_{10}\alpha_{\rmn{CE}}=\epsilon_{0},
\label{const}
\end{equation}
where $\epsilon_{0}$ is a constant. For the data in
Fig. \ref{ace_params_G}, we find that $\epsilon_{0}=-0.06\pm 0.04$.

Values of $\mathcal{F}$ [and the probability that we exceed this value,
$\mathscr{P}(\mathcal{F})$] for the data in Fig. \ref{ace_params_G} are
given in Table \ref{fits_g}. Table \ref{fits_g} also indicates whether we
can reject the null hypothesis $\mathscr{H}^{\mathcal{F}}_{0}$ (see Section
\ref{subsection:F_test}).

\subsubsection{A fit in terms of the progenitor parameters}

Table \ref{fits_g} indicates that we can reject
$\mathscr{H}_{0}^{\mathcal{F}}$ for $P_{CE,i}$ and $M_{2}$. The first
variable which we add to our fit is $P_{\rmn{CE,i}}$ as this has the
smallest value of $\mathscr{P}(\mathcal{F})$. Subsequent variables which we
add, and the corresponding fit parameters, are summarised in the top panel
of Table \ref{regress_g}. Specifically, Table \ref{regress_g} gives the
values $\beta_{0...m}$, the F-statistic $\mathcal{F}$ and
$\mathscr{P}(\mathcal{F})$. We also indicate whether we can reject
$\mathscr{H}_{0}^{\mathcal{F}}$ for each fit. We adopt a significance level
of 0.05.

We find that adding binary parameters in addition to $P_{\rmn{CE,i}}$ do
not provide an improved fit over one in terms of $P_{\rmn{CE,i}}$
only. Indeed, $\mathscr{P}(\mathcal{F})>0.05$ in all cases. For such fits,
we therefore cannot reject $\mathscr{H}_{0}^{\mathcal{F}}$. Hence, we find
the following fit of $\alpha_{\rmn{CE}}$ in terms of $P_{\rmn{CE,i}}$ which
is given by
\begin{eqnarray}
  \rmn{log}_{10}\alpha_{\rmn{CE}}=(0.94\pm 0.33)-(0.34\pm
  0.11)\rmn{log}_{10}\left(\frac{P_{\rmn{CE,i}}}{\rmn{d}}\right).
\label{progenitor_model_G}
\end{eqnarray}
Our reconstructed values of $\alpha_{\rmn{CE}}$ versus values calculated
from equation (\ref{progenitor_model_G}) are shown in
Fig. \ref{fig_progenitor_model_G}.

\subsubsection{A fit in terms of the PCEB parameters}

The first variable which we add to our fit is $P_{\rmn{CE}}$. Subsequent
variables that we add, and the corresponding fit parameters, are summarised
in the bottom panel of Table \ref{regress_g}. In contrast to our fit of
$\alpha_{\rmn{CE}}$ in terms of the progenitor parameters, we find that
adding $M_{\rmn{WD}}$ and then $M_{2}$ is an improvement over the fit which
is only in terms of $P_{\rmn{CE}}$; we have $\mathcal{F}=16.6$ and 38.4 for
the former and latter cases respectively. Hence, we can reject
$\mathscr{H}_{0}^{\mathcal{F}}$ for both of these alternative fits. Hence,
we suggest
\begin{eqnarray}
\rmn{log}_{10}{\alpha_{\rmn{CE}}}=-0.31-(0.95\pm
0.13)\rmn{log}_{10}\left(\frac{M_{\rmn{WD}}}{\rmn{M}_{\odot}}\right) \nonumber \\ +(0.28\pm
0.02)\rmn{log}_{10}\left(\frac{P_{\rmn{CE}}}{\rmn{d}}\right) \nonumber \\
-(0.17 \pm 0.09)\rmn{log}_{10}\left(\frac{M_{2}}{\rmn{M}_{\odot}}\right).
\label{PCEB_model_G}
\end{eqnarray}
Fig. \ref{fig_PCEB_model_G} is the same as
Fig. \ref{fig_progenitor_model_G} except equation (\ref{PCEB_model_G}) is
used.

\subsection{For $\lambda=\lambda_{\rmn{b}}$}

\begin{figure*}
  \begin{minipage}{175mm}
  \begin{center}
    \includegraphics[scale=0.80]{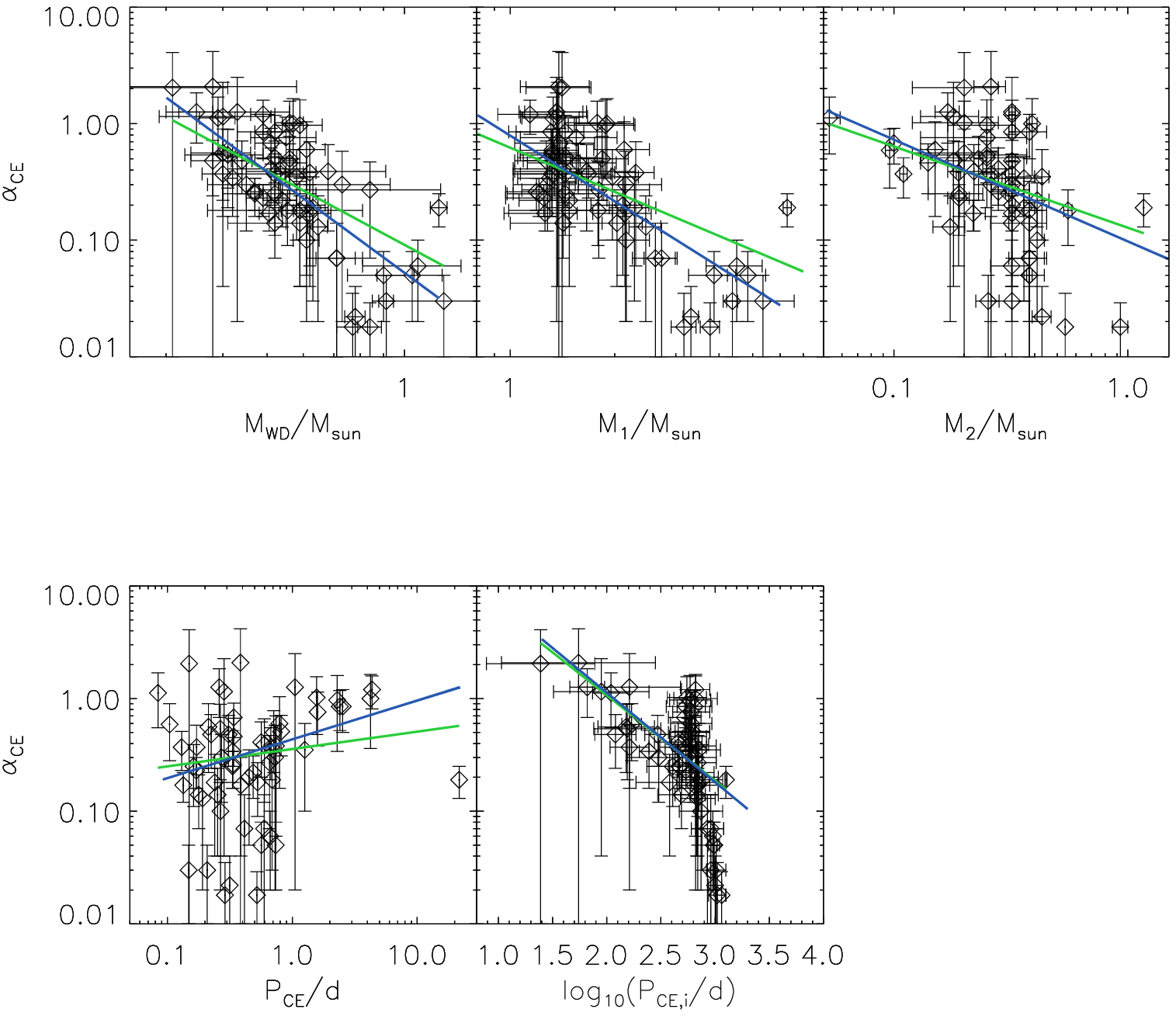}
    \caption{Similar to Fig. \ref{ace_params_G} except for the case where
      we take $\lambda=\lambda_{\rmn{b}}$. The solid blue lines indicate
      linear fits to the data using the bootstrapping technique (see
      Section \ref{section:discussion} for further details).}
    \label{ace_params_B}
  \end{center}
  \end{minipage}
\end{figure*}

The reconstructed values of $\alpha_{\rmn{CE}}$ versus each binary
parameter are shown in the corresponding panels in Fig.
\ref{ace_params_B}. The values of $\epsilon_{0}$, $\epsilon_{1}$ are
summarised in Table \ref{fits_b}, as are the $\mathcal{F}$-statistics, and
whether we can reject $\mathscr{H}_{0}^{\mathcal{F}}$. For comparison with
our linear fits, we once again perform a constant fit to our values of
$\alpha_{\rmn{CE}}$ in the form of eqn. (\ref{const}).We find that
$\epsilon_{0}=-0.49\pm 0.05$.

\subsubsection{A fit in terms of the progenitor parameters}

Table \ref{fits_b} indicates that we can reject
$\mathscr{H}_{0}^{\mathcal{F}}$ for the 1D fit of $\alpha_{\rmn{CE}}$
versus $M_{\rmn{WD}}$, $M_{1}$, $M_2$ or $P_{\rmn{CE,i}}$. We obtain the
smallest value of $\mathscr{P}(\mathcal{F})$ for $M_{\rmn{WD}}$, and so
this is the first binary parameter we add to our fit. Subsequent additions
and the corresponding fit parameters are summarised in the top panel of
Table \ref{regress_b}, which shows the same quantities as displayed in
Table \ref{regress_g}.

Adding more binary parameters to our fit does not provide an improvement
over the fit only in terms of $M_{\rmn{WD}}$. This is probably a result of
the fact that trends exist between $M_{1}$, $M_{\rmn{WD}}$ and
$P_{\rmn{CE,i}}$. Therefore, we suggest a fit of $\alpha_{\rmn{CE}}$ as a
function of $M_{\rmn{WD}}$ only, i.e.
\begin{equation}
\rmn{log}_{10}\alpha_{\rmn{CE}}=(-1.03\pm 0.10) -(2.10\pm
0.34)\rmn{log}_{10}\left(\frac{M_{\rmn{WD}}}{\rmn{M}_{\odot}}\right).
\label{progenitor_model_B}
\end{equation}
Fig. \ref{fig_progenitor_model_B} is the same as
Fig. \ref{fig_progenitor_model_G} except equation
(\ref{progenitor_model_B}) is used.

\subsubsection{A fit in terms of the PCEB parameters}

\begin{figure}
  \begin{center}
    \includegraphics[scale=0.5]{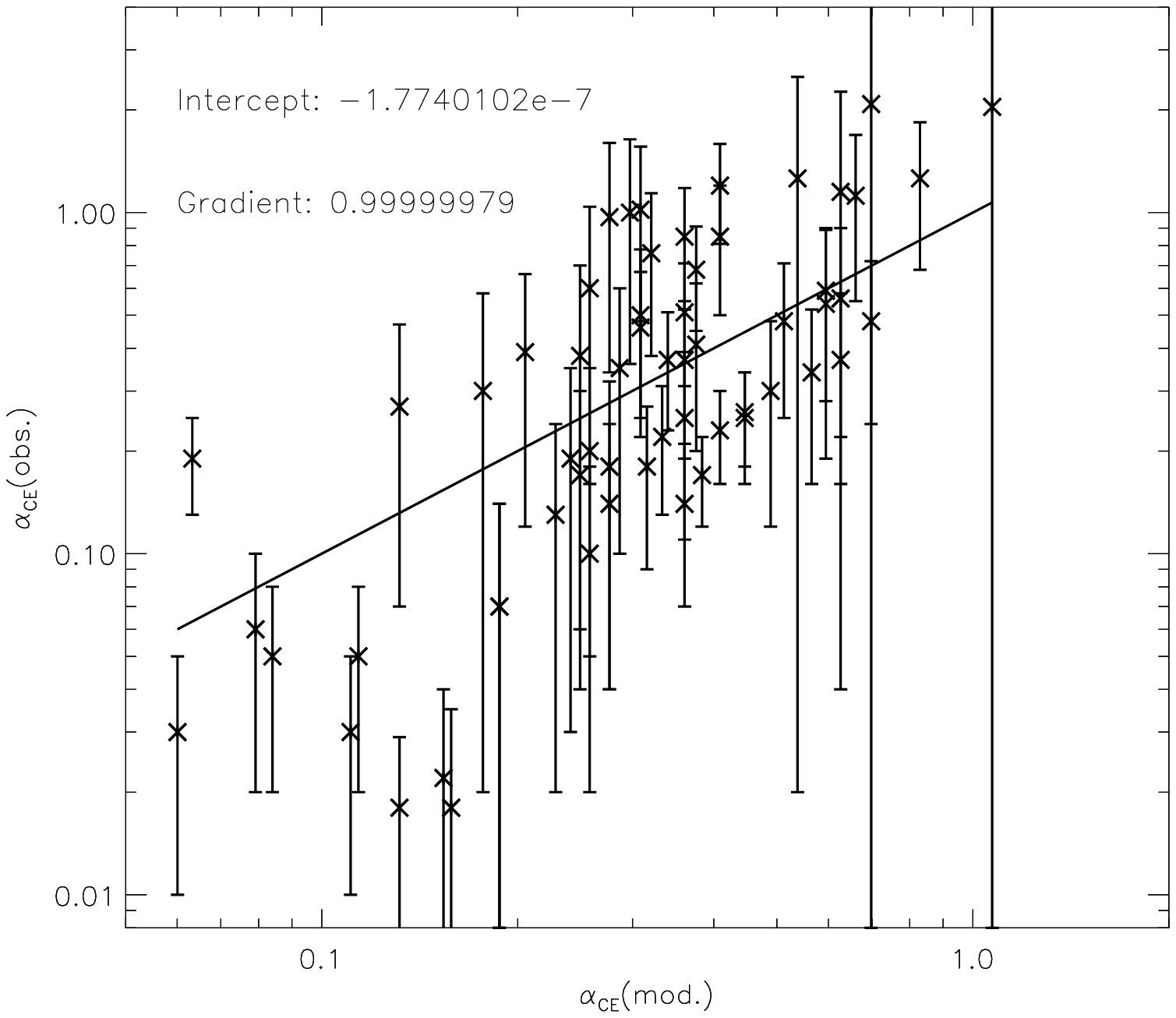}
    \caption{Same as Fig. \ref{fig_progenitor_model_G}, except
      $\alpha_{\rmn{CE}}(\rmn{mod.})$ has been calculated from
      eqn. (\ref{progenitor_model_B}).}
    \label{fig_progenitor_model_B}
    \includegraphics[scale=0.5]{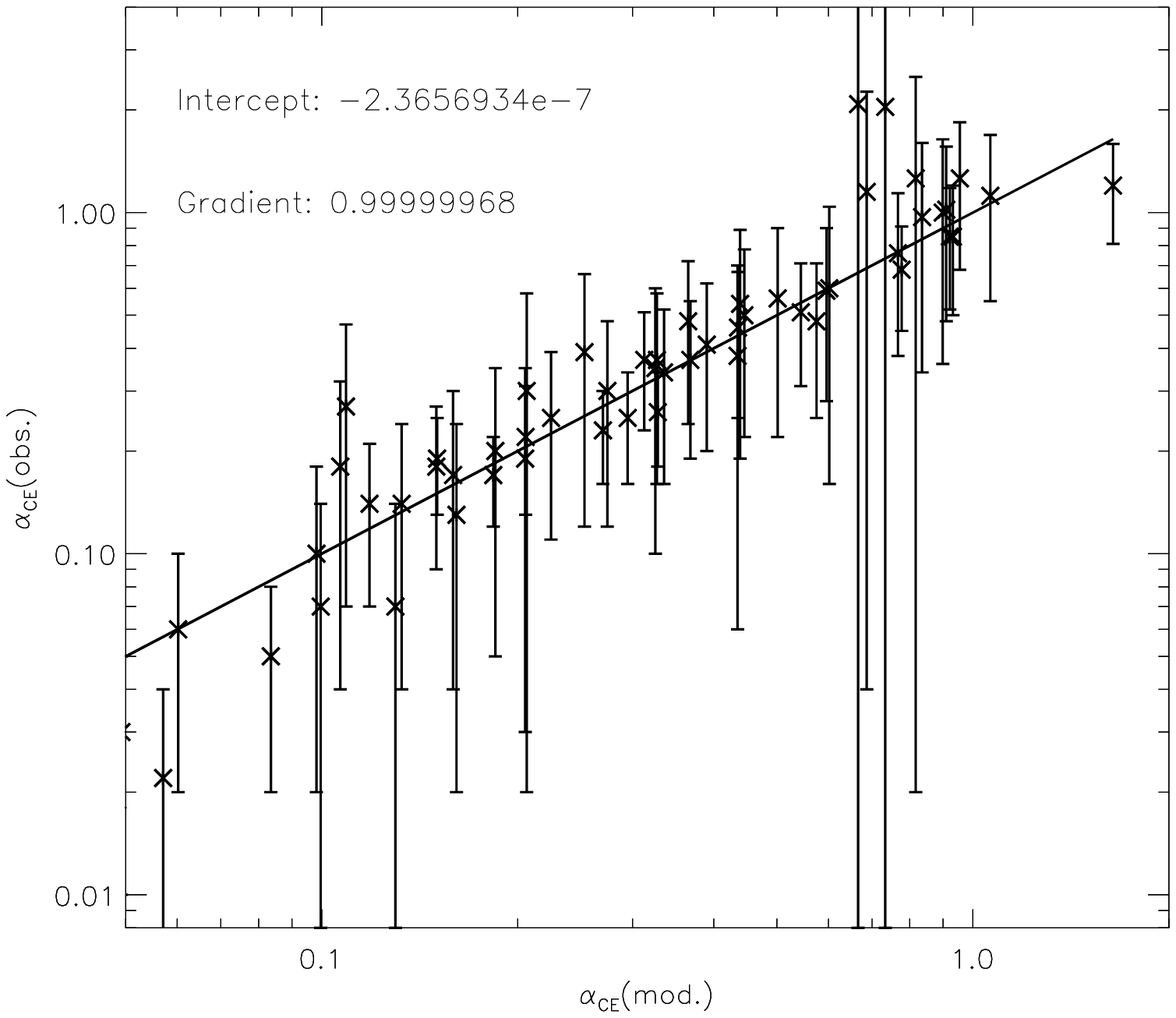}
    \caption{Same as Fig. \ref{fig_progenitor_model_G}, except
      $\alpha_{\rmn{CE}}(\rmn{mod.})$ has been calculated from
      eqn. (\ref{PCEB_model_B}).}
    \label{fig_PCEB_model_B}
  \end{center}
\end{figure}

We once again start with $M_{\rmn{WD}}$ in our fit, with the fit parameters
for additional variables summarised in the bottom panel of Table
\ref{regress_b}. Including $P_{\rmn{CE}}$ and then $M_{2}$ provides an
improved fit over one in terms of $M_{\rmn{WD}}$ only; we can reject
$\mathscr{H}_{0}^{\mathcal{F}}$ in either case. As for the case where
$\lambda=\lambda_{\rmn{g}}$, we suggest a fit of $\alpha_{\rmn{CE}}$ as a
function of $M_{\rmn{WD}}$, $M_{2}$ and $P_{\rmn{CE}}$, given by
\begin{eqnarray}
 \rmn{log}_{10}{\alpha_{\rmn{CE}}}=-1.18-(2.50\pm
0.16)\rmn{log}_{10}\left(\frac{M_{\rmn{WD}}}{\rmn{M}_{\odot}}\right) \nonumber \\ +(0.31\pm
0.02)\rmn{log}_{10}\left(\frac{P_{\rmn{CE}}}{\rmn{d}}\right) \nonumber \\
-(1.02 \pm 0.09)\rmn{log}_{10}\left(\frac{M_{2}}{\rmn{M}_{\odot}}\right).
\label{PCEB_model_B}
\end{eqnarray} 
Fig. \ref{fig_PCEB_model_B} is the same as
Fig. \ref{fig_progenitor_model_G} except equation (\ref{PCEB_model_B}) is
used.

\section{Discussion}
\label{section:discussion}

We have found statistical evidence that the CE ejection efficiency has at
least a first order dependence on the white dwarf mass, the progenitor
primary mass, the secondary mass or the orbital period at the point the CE
phase commences if we consider the case where
$\lambda=\lambda_{\rmn{b}}$. For $\lambda=\lambda_{\rmn{g}}$, we find
correlations between $\alpha_{\rmn{CE}}$ and $M_{2}$, $P_{\rmn{CE}}$ or
$P_{\rmn{CE,i}}$. However, whether we consider $\lambda=\lambda_{\rmn{g}}$
or $\lambda=\lambda_{\rmn{b}}$, we do not obtain a statistically
significant fit of $\alpha_{\rmn{CE}}$ in terms of more than one progenitor
binary parameter.

This behaviour may be a result of the large uncertainties of the progenitor
system parameters for a given observed PCEB. As Fig. \ref{examples}
demonstrates, an uncertainty for the white dwarf mass of a few$\times 0.01$
M$_{\odot}$ can translate into possible progenitor primary masses that span
between approximately 1 and 2 M$_{\odot}$, or into a pre-CE orbital period
that can range from a few hundred to approximately 1000 d. As Tables
\ref{CERecon_SDSS} to \ref{CERecon_RKCat_B} show, this can result in a
standard deviation in $\alpha_{\rmn{CE}}$ as large as approximately 100 per
cent of the mean value. This presents a challenge to efforts aimed at
discerning potentially small, albeit real, trends with binary parameters.

Even if the trends which we see are statistically significant, it is
unclear whether they have a physical underpinning. The initial
distribution functions which we use to weight the possible values of
$\alpha_{\rmn{CE}}$ for a given PCEB have been inferred from binary
populations which may be prone to selection effects. Hence, for a
given observed system, the calculated average value of
$\alpha_{\rmn{CE}}$ may not be the true average
$\alpha_{\rmn{CE}}$. 

We now address whether there is a source of energy in addition to
gravitational energy which is used during the ejection of the CE. We find
that if we do consider the internal energy of the progenitor primary
envelope, then this does bring the ejection efficiencies within the
canonical range of $0<\alpha_{\rmn{CE}}\leq{1}$, in agreement with
\citet{Zorotovic2010} and as predicted by \citet{Webbink2008}. However, we
still find that $\alpha_{\rmn{CE}}\ga{1}$ for PCEBs with brown dwarf
secondaries ($M_{2}\leq{0.1}$ M$_{\odot}$).

There is some debate as to what extent the internal energy of the
giant's envelope plays a role in the ejection of the CE. Indeed,
\citet{Harpaz1998} argues that during the planetary nebula phase, the
opacity of the giant's hydrogen ionization zone decreases during
recombination. Hence, the envelope becomes transparent to its own
radiation. The radiation will therefore freely escape rather than push
against the material to eject it. On the other hand,
\citet{Webbink2008} argues that the ionization zone of giant and AGB
stars is buried beneath a region of the envelope whose opacity is
dominated by heavy elements. Once recombination is triggered at the
onset of the CE phase, the escaping radiation can therefore `push'
against these high-opacity layers.

\citet{Ivanova2011b}, on the other, suggest that the enthalpy of the
primary giant's envelope, as opposed to the internal energy of the gas,
should be considered. Indeed, they found that there exists a region in the
giant interior, coined the `boiling pot zone', where the enthalpy is
greater than the gravitational potential. This excess of energy may cause
an outflow of material lying above the boiling-pot zone, and hence
facilitate the ejection of the CE from the system. They find that by
considering the enthalpy of the progenitor giant permits the formation of
low-mass X-ray binaries with a low-mass companion, without resorting to
$\alpha_{\rmn{CE}}>1$.

Our calculations with $\lambda=\lambda_{\rmn{b}}$ assume that all of
the internal energy of the envelope is used to unbind the CE from the
system, i.e $\alpha_{\rmn{th}}=1$ in eqn. (\ref{Ebind_b}), which is
clearly unrealistic. Nonetheless, by reconstructing the CE phase with
and without consideration for the internal energy of the primary
envelope, we provide upper and lower limits for
$\alpha_{\rmn{CE}}$. Furthermore, our reconstructed values of
$\alpha_{\rmn{CE}}$ and their trends with binary parameters presented
in Section \ref{section:results} will provide useful constraints on
detailed models of the CE microphysics.

Another result to note from our investigation regards the progenitors of
those PCEBs with high mass white dwarfs. For PCEBs with white dwarf masses
$M_{\rmn{WD}}\ga 0.7$ M$_{\odot}$, there are possible progenitor primaries
which have $\lambda_{\rmn{b}}<0$ at the onset of the CE phase (see also
Section \ref{subsection:BiSEPS}). The envelope is therefore unbound from
the giant as a result of radiation pressure driven winds. A particularly
striking example is IK Peg, where a vast majority of the possible
progenitors have $\lambda_{\rmn{b}}<0$ (see Fig. \ref{HighMwd}). This begs
the question: did IK Peg really form via a CE phase? \citet{Dermine2009}
investigated the effect that radiation pressure from giant stars has on the
Roche lobe geometry. If $f$ is the ratio of the radiation to the
gravitational force, they found that the critical Roche lobe around the
donor star becomes meaningless if $f=1$. Hence, mass lost from the star
will be in the wind regime, and material may not necessarily be channelled
directly into the Roche lobe of the secondary star. Hence, in the context
of giant stars, a situation where $f \geq 1$ may mean that a CE phase is
avoided. Theoretical studies investigating this possibility need to be
carried out further.

In the following Sections, we discuss important points regarding the
reconstructed values of $\alpha_{\rmn{CE}}$ versus each binary parameter in
turn.

\subsection{$\alpha_{\rmn{CE}}$ versus $M_{\rmn{WD}}$}
\label{subsection:alpha_v_Mwd}

The following argument may further help to explain why we can reject more
frequently $\mathscr{H}^{\mathcal{F}}_{0}$ for the data calculated using
$\lambda=\lambda_{\rmn{b}}$ rather than when $\lambda=\lambda_{\rmn{g}}$ is
used. Recall that there are some PCEB progenitor primaries with
$M_{\rmn{WD}}\ga{0.7}$ M$_{\odot}$ which have $\lambda_{\rmn{b}}<0$ (see
Section \ref{section:discussion}). Furthermore, the progenitor primaries of
these PCEBs which \emph{do} have $\lambda_{\rmn{b}}>0$ have very large
values of $\lambda_{\rmn{b}}$ (and a correspondingly small binding
energies) owing to their large radii when they fill their Roche lobes at
the start of the CE phase. As a result, $\alpha_{\rmn{CE}}$ will be very
small (for example IK Peg, where we find a mean CE ejection efficiency of
$\alpha_{\rmn{CE}}=0.19$). This explains why the values of $|\epsilon_{1}|$
are larger when we consider $\lambda=\lambda_{\rmn{b}}$ than when we
consider $\lambda=\lambda_{\rmn{g}}$.

\begin{figure}
  \begin{center}
  \includegraphics[scale=0.4]{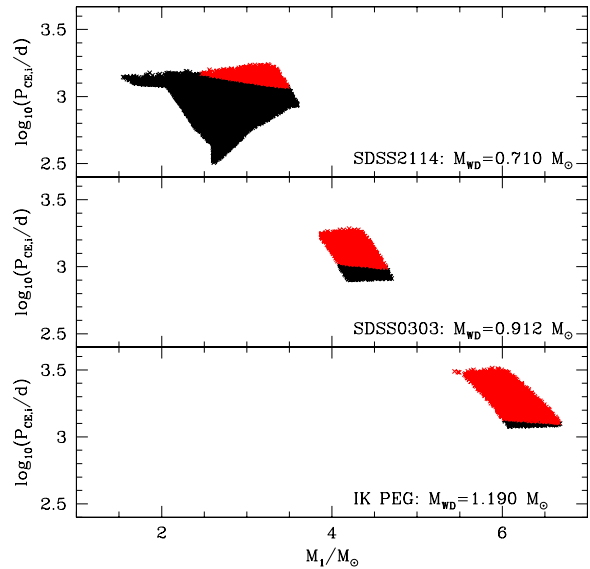}
  \caption{Two dimensional distributions in primary mass, $M_{1}$ and
    orbital period, $P_{\rmn{CE,i}}$, at the very start of the CE
    phase, for three systems as indicated in the bottom right of each
    panel.Black points indicate that $\lambda_{\rmn{b}}>0$ for those
    progenitors, while red data points indicate that
    $\lambda_{\rmn{b}}<0$.}
  \label{HighMwd}
  \end{center}
\end{figure}

Furthermore, we find that as the white dwarf mass increases, the number of
possible progenitor primaries with $\lambda_{\rmn{b}}>0$ decreases. This is
graphically illustrated in Fig. \ref{HighMwd}, which shows the
two-dimensional distribution in $M_{1}$ and $P_{\rmn{CE,i}}$ for three
systems: SDSS2214-0103, SDSS0303-0054 and IK Peg. Black points indicate
that $\lambda_{\rmn{b}}>0$ for the associated progenitor, while red points
indicate that $\lambda_{\rmn{b}}<0$ for that progenitor. 

This means that the range of possible progenitor primary masses, and
orbital periods at the start of the CE phase, and consequently the range in
$\alpha_{\rmn{CE}}$ will decrease. As a result, the standard deviation in
$\alpha_{\rmn{CE}}$ will decrease. Such data points will therefore have a
large weighting in the fits. The linear fits between $\alpha_{\rmn{CE}}$
and $M_{\rmn{WD}}$ may be greatly influenced by such data points.

To determine if this is the case, and to determine if a linear relationship
does exist among the other data points, we perform a bootstrap analysis to
the $\alpha_{\rmn{CE}}$ versus $M_{\rmn{WD}}$ data in
Fig. \ref{ace_params_B}. The mean value of $\epsilon_{1}$ and the standard
error from this analysis is given in Table \ref{bootstrap_params}. The
linear fit calculated from our bootstrapping analysis are shown in
Fig. \ref{ace_params_B} as solid blue curves.

\begin{table}
  \caption{The mean values with errors of the coefficient
    $\epsilon_{1}$ calculated from our bootstrapping analysis of the
    data in each panel of Fig. \ref{ace_params_B}.}
  \begin{center}
  \begin{tabular}{@{}lc@{}}
    \hline
    $Q$     &     $\epsilon_{1}$  \\
    \hline
    $M_{\rmn{WD}}$       &   -2.88$\pm$0.33     \\ 
    $M_{1}$             &   -1.87$\pm$0.28    \\
    $M_{2}$             &   -0.87$\pm$0.24    \\
    $P_{\rmn{CE}}$       &    0.34$\pm$0.11    \\
    $P_{\rmn{CE,i}}$     &   -0.79$\pm$0.15     \\
    \hline
  \end{tabular}
  \label{bootstrap_params}
  \end{center}
\end{table}

In comparison to the linear fit, we find that the bootstrap analysis also
indicates that $\alpha_{\rmn{CE}}$ decreases with increasing
$M_{\rmn{WD}}$, albeit a slightly steeper (i.e. a more negative) value of
$\epsilon_{1}$ is predicted. In other words, a marginally stronger linear
relationship between $\alpha_{\rmn{CE}}$ and $M_{\rmn{WD}}$ is predicted by
the bootstrapping analysis. By comparing the green and blue lines in
Fig. \ref{ace_params_B} we can see that the data points at large white
dwarf masses have the effect of flattening the gradient. The majority of
the data points, however, suggest a steeper gradient.

\subsection{$\alpha_{\rmn{CE}}$ versus $M_{1}$}
\label{subsection:alpha_v_M1}

The linear fits to the corresponding panels in Fig. \ref{ace_params_G}
and \ref{ace_params_B} suggest that $\alpha_{\rmn{CE}}$ decreases with
increasing $M_{1}$. This is in contrast to \citet{DeMarco2011}, who
found that $\alpha_{\rmn{CE}}$ increases with $M_{1}$. We explore this
further in Section \ref{subsection:comparison_DM}.

In light of the influence that data points corresponding to PCEBs with high
mass white dwarfs may have on the fitting (see Section
\ref{subsection:alpha_v_Mwd}) we also perform a bootstrapping analysis on
the $M_{1}$ versus $\alpha_{\rmn{CE}}$ data in Fig. \ref{ace_params_B}. (We
repeat this analysis for the subsequent binary parameters in our
discussion). Our bootstrap analysis predicts a slightly more negative value
than the value obtained from the chi-squared fit. Therefore, those data
points corresponding to high mass white dwarfs do not significantly affect
the linear fit.

\subsection{$\alpha_{\rmn{CE}}$ versus $M_{2}$}
\label{subsection:alpha_v_M2}

We find that $\alpha_{\rmn{CE}}$ decreases with increasing $M_{2}$,
irrespective of whether we use $\lambda_{\rmn{g}}$ or
$\lambda_{\rmn{b}}$. This finding is in agreement with \citet{DeMarco2011},
but in contrast to the suggestion made by \citet{Politano2007}, who
proposed that $\alpha_{\rmn{CE}}$ may increase with increasing $M_{2}$.

This suggestion was motivated by the fact that very few, if any, CVs with
brown dwarf secondaries at orbital periods below 77 min have been
detected. This appeared to be in conflict with \citet{Politano2004} who
estimated from his models that such systems should make up approximately 15
per cent of the total CV population \citep[see
also][]{Kolb1999}. \citet{Politano2004} suggested that this discrepancy may
be a result of the decreasing energy dissipation rate of orbital energy
within the CE for decreasing secondary mass, and that below some cut-off
mass, a CE merger would be unavoidable. Indeed, the small surface area of
the secondary star will generate a correspondingly small friction force
with the CE material. Also, as can be seen from eqn. (\ref{Eorb_5}), the
change in orbital energy during spiral-in will be small for small secondary
masses.

However, as PCEBs with $M_{2}\la{0.1}$ M$_{\odot}$ are observed, the CE
ejection must be very efficient if such systems are to avoid a
merger. \citet{DeMarco2011} suggested this may be achieved if smaller mass
components take longer than the giant's dynamical time to penetrate into
the envelope. The giant star would therefore have time to thermally
re-adjust its structure, and use this thermal energy to aid the ejection of
the CE. Note, however, that systems with secondaries $M_{2}\approx{0.1}$
M$_{\odot}$ (see Fig. \ref{ace_params_B}), require a mean ejection
efficiency close to 1, even if we consider the internal energy of the
progenitor primary's envelope.

The bootstrap analysis yields only a marginally steeper gradient than that
obtained from the chi-squared fit.

\subsection{$\alpha_{\rmn{CE}}$ versus $P_{\rmn{CE}}$}
\label{subsection:alpha_v_Pce}

We find that $\alpha_{\rmn{CE}}$ increases with increasing $P_{\rmn{CE}}$,
irrespective of whether we consider the internal energy of the progenitor
primaries or not.

Furthermore, the orbital separation at the point the CE terminates,
$A_{\rmn{CE,f}}$ (found via $P_{\rmn{CE}}$), for the observed PCEBs
indicates how deep the secondary star penetrated into the envelope of
the giant star by the time the CE ceased. This may shed further light
into the role that the structure of the giant primary plays during the
termination of the CE phase. \citet{Yorke95} suggested that a steep
decrease in density within the giant star is required for the
successful termination of the CE phase. Such a decrease is
characterized by a flat mass-radius profile of the star.

Once the secondary star reaches this region, the energy dissipation rate
within the CE decreases, and the CE phase comes to an end, provided that
the envelope material exterior to this region has been unbound from the
system. \citet{Terman1996} suggest that the values of $A_{\rmn{CE,f}}$ for
PCEBs may be estimated from the location within the CE where the
mass-radius profile is no longer flat, i.e. when the quantity
$V=\rmn{d}\,\rmn{ln}P/\rmn{d}\,\rmn{ln}r$ is at a minimum. Here, $P$ is the
pressure and $r$ is the radial coordinate.

\citet{Terman1996} found from their hydrodynamical simulations that
the orbital separation of PCEBs immediately after the CE phase was
proportional to the white dwarf mass (see their Fig. 8), as a
consequence of the fact that the point where $V$ is minimum moves
further away from the stellar centre for more evolved
stars. Therefore, a plot of $M_{\rmn{WD}}$ versus $A_{\rmn{CE,f}}$ for
our observed sample of PCEBs would provide an excellent comparison
with the aforementioned theoretical study.

\begin{figure}
  \begin{center}
    \includegraphics[scale=0.55]{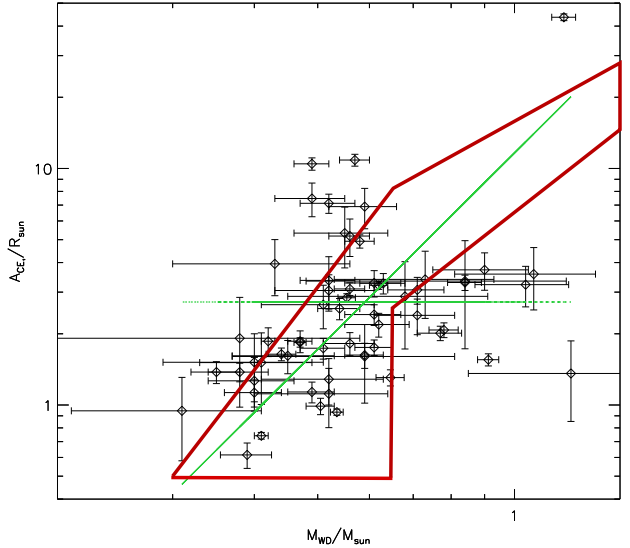}
    \caption{A plot of white dwarf mass versus the post-CE orbital
      separation for our observed sample of PCEBs. The green dotted line is
      a constant fit to the data, while the green solid line is a linear
      fit to the data. The thick red border outlines the approximate
      region in $A_{\rmn{CE,f}}-M_{\rmn{WD}}$ space inhabited by the data
      points given in Fig. 8 of \citet{Terman1996}, calculated from their
      hydrodynamical simulations.}
    \label{Mwd_Asepf}
  \end{center}
\end{figure}

Such a plot is shown in Fig. \ref{Mwd_Asepf}. To determine if a linear
relationship between $M_{\rmn{WD}}$ and $A_{\rmn{CE,f}}$ does exist, we fit
a linear function given by
\begin{equation}
  \rmn{log}_{10}A_{\rmn{CE,f}}=\epsilon_{0}+\epsilon_{1}\rmn{log}_{10}M_{\rmn{WD}},
\label{Asepf_Mwd_linr}
\end{equation}
where $A_{\rmn{CE,f}}$ is expressed in solar radii, and $M_{\rmn{WD}}$ is
expressed in solar masses. We find $\epsilon_{0}=1.07\pm 0.01$ and
$\epsilon_{1}=2.76\pm 0.03$. For our value of the number of degrees of
freedom, $\nu_{\rmn{linr}}=58$, we therefore obtain a reduced chi-squared
value of $\chi^{2}_{\rmn{linr}}=98.62$. This curve is shown as the green
solid line in Fig. \ref{Mwd_Asepf}. For comparison, we also show a constant
fit to the data, $\rmn{log}_{10}A_{\rmn{CE,f}}=\epsilon_{0}$, where
$\epsilon_{0}=0.435\pm 0.004$, and $\chi^{2}_{\rmn{const}}=213.39$
(reduced), for $\nu_{\rmn{const}}=59$ degrees of freedom.

The chi-squared values indicate that a linear function provides a better
fit to the data in Fig. \ref{Mwd_Asepf} than a constant one. To determine
whether an extra parameter is warranted, we once again apply the
F-Test. For our values of $\nu_{\rmn{linr}}$ and $\nu_{\rmn{cons}}$, we
obtain $\mathcal{F}=67.5$ with $\mathscr{P}(\mathcal{F})<10^{-6}$. Hence,
because $\mathscr{P}(\mathcal{F})<\alpha$, we can reject
$\mathscr{H}^{\mathcal{F}}_{0}$. We therefore find intriguing statistical
evidence that there is a functional dependence between $M_{\rmn{WD}}$ and
$A_{\rmn{CE,f}}$.

The thick red border in Fig. \ref{Mwd_Asepf} indicates the region in the
$A_{\rmn{CE,f}}-M_{\rmn{WD}}$ plane inhabited by the data points in Fig. 8
of \citet{Terman1996}, calculated from their hydrodynamical simulations. It
can be seen that there is reasonable overlap between the observed PCEB
sample and the results of \citet{Terman1996} for $M_{\rmn{WD}}\la{0.7}$
M$_{\odot}$. In contrast, their simulations over-estimate the final orbital
separation by as large as a factor of about 3 for systems with more massive
white dwarf primaries.

However, as noted by \citet{Yorke95}, the orbital separation when the
CE terminates does not necessarily coincide exactly with the point
when $V$ becomes a minimum. \citet{Yorke95} and \citet{Terman1996}
estimate that the orbital separation at which the orbital decay
sufficiently decreases is approximately 3 to 10 times smaller than the
distance when $V$ becomes a minimum. \citet{Terman1996} assume a
factor of one sixth of this distance in their Fig. 8, although it is
more likely that this factor will differ from system to
system. 

More recently, \citet{Ivanova2011} discussed the idea of the `compression
point' within the primary giant, and the role that this may play during the
CE ejection phase. The compression point is the location within the giant
primary star where the value $P/\rho$ (here $P$ is the pressure and $\rho$
is the density) is at a maximum. \citet{Ivanova2011} found that, if the
companion star is sufficiently massive, the envelope will be unbound up to
the mass coordinate corresponding to the location of the compression
point. Hence, the final orbital separation at the end of the CE phase may
correspond with the radial coordinate associated with the compression
point. Hence, our fit here may provide a useful diagnostic of the
termination of the CE phase.

\subsection{$\alpha_{\rmn{CE}}$ versus $P_{\rmn{CE,i}}$}
\label{subsection:alpha_v_Pcei}

The data points at high values of log$_{10}(P_{\rmn{CE,i}}/\rmn{d})$
in Fig. \ref{ace_params_B} may overly influence the linear fits. The
reason is the same as for those data points at high white dwarf
masses, as discussed in Section \ref{subsection:alpha_v_Mwd}:
primaries that fill their Roche lobes towards longer orbital periods
will have very large or negative values of $\lambda_{\rmn{b}}$. Fewer
progenitor systems will therefore have $\lambda_{\rmn{b}}>0$,
decreasing the range in possible values of $\alpha_{\rmn{CE}}$.

We also find that the difference between $\epsilon_{1}$ calculated from the
chi-squared fit and our bootstrap calculation is negligible; our bootstrap
analysis gives a slightly steeper value of $\epsilon_{1}$ compared with
that obtained from the chi-squared fit.

\begin{figure}
  \begin{center}
    \includegraphics[scale=0.40]{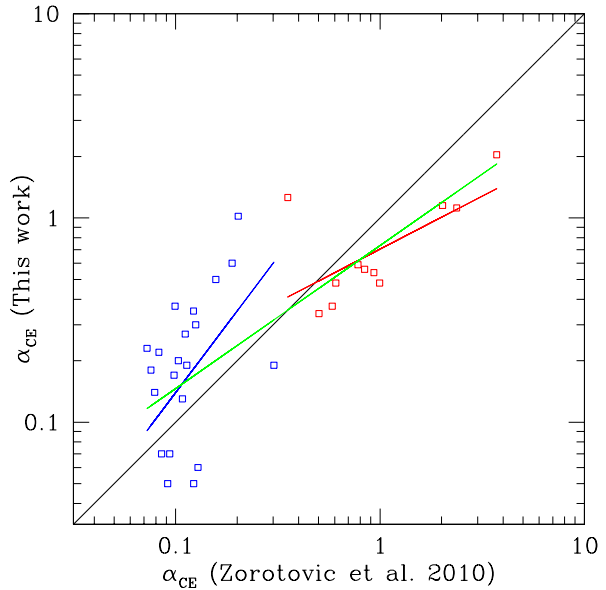}
    \caption{A comparison of our reconstructed values of
      $\alpha_{\rmn{CE}}$ (y-axis) with those calculated by
      \citet{Zorotovic2010} (x-axis). PCEBs formed only from a primary
      progenitor on the first giant branch are indicated by the red points,
      while blue points indicate that the progenitor primary was on the
      AGB. The solid black line is of the form $\alpha_{\rmn{CE}}$(this
      work)$=$$\alpha_{\rmn{CE}}$\citep{Zorotovic2010}. Fits to the red and
      blue data points are shown as the red and blue lines respectively,
      while the green solid line is a linear fit to all the data
      points. Here, and in Figs. \ref{compare_DM} and \ref{ace_M1_M2_Q},
      error-bars are omitted for clarity.}
    \label{compare_zorotovic}
  \end{center}
\end{figure}

\begin{figure}
  \begin{center}
    \includegraphics[scale=0.40]{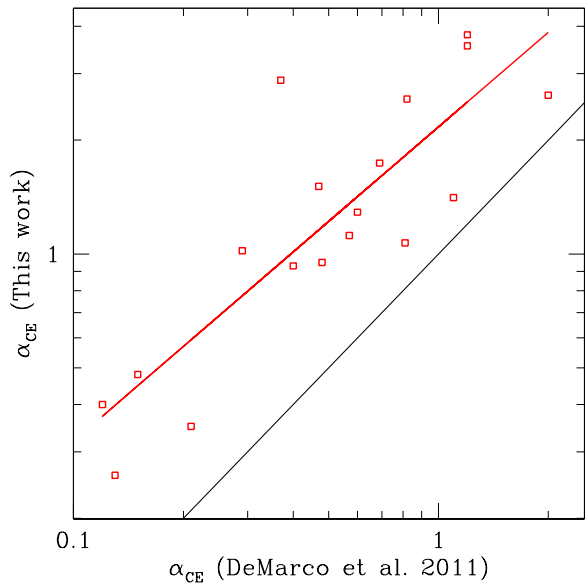}
    \caption{Similar to Fig. \ref{compare_zorotovic} except here we compare
      our values of $\alpha_{\rmn{CE}}$ with those calculated by
      \citet{DeMarco2011} (red data points).}
    \label{compare_DM}
  \end{center}
\end{figure}

\section{Comparisons with previous studies}
\label{section:comparisons}

\subsection{Comparisons with \citet{Zorotovic2010}}
\label{subsection:comparison_zorotovic}

Fig. \ref{compare_zorotovic} compares our values of $\alpha_{\rmn{CE}}$
calculated with $\lambda=\lambda_{\rmn{b}}$ (y-axis) with those values
calculated by \citet{Zorotovic2010} (x-axis). The black line indicates
$y=x$.

For a given PCEB, \citet{Zorotovic2010} determine mean values of
$\alpha_{\rmn{CE}}$ separately from progenitors that commenced the CE phase
on either the first giant branch (FGB) or on the AGB. Hence,
\citet{Zorotovic2010} provide two mean values of $\alpha_{\rmn{CE}}$ for
some PCEBs. This is in contrast to the present investigation, where we
calculate one mean value of $\alpha_{\rmn{CE}}$ for each PCEB, from all
possible progenitors.

Therefore, to place our comparison in Fig. \ref{compare_zorotovic} on an
equal footing, we only compare values of $\alpha_{\rmn{CE}}$ for systems
that only have FGB progenitors (shown as red points in
Fig. \ref{compare_zorotovic}) and for systems with only AGB progenitors
(blue points). The red and blue solid lines are linear fits to the
corresponding data points, while the green curve is a fit to all the data
points.

For the case of those systems which have AGB progenitors, we typically
obtain larger values of $\alpha_{\rmn{CE}}$ than \citet{Zorotovic2010}, by
no more than a factor of 2. For PCEBs with FGB progenitors, we typically
get values of $\alpha_{\rmn{CE}}$ which are smaller by a factor of no more
than about 3.

While in this present work we always take $\alpha_{\rmn{th}}=1$,
\citet{Zorotovic2010} assume that the efficiency of using the giant
envelope's internal energy to eject the CE is equal to the efficiency of
using the orbital energy. Their values of $\lambda$ are then calculated
from this assumption. Furthermore, \citet{Zorotovic2010} calculate their
values of $\lambda$ using analytical fits to the values of
$E_{\rmn{bind}}$, calculated from the stellar models of
\citet{Pols1998}. This is in contrast to the tabulated values of
$\lambda_{\rmn{b}}$ for $\alpha_{\rmn{th}}=1$ used in the present
investigation.

\begin{figure*}
  \begin{center}
    \includegraphics[scale=0.35]{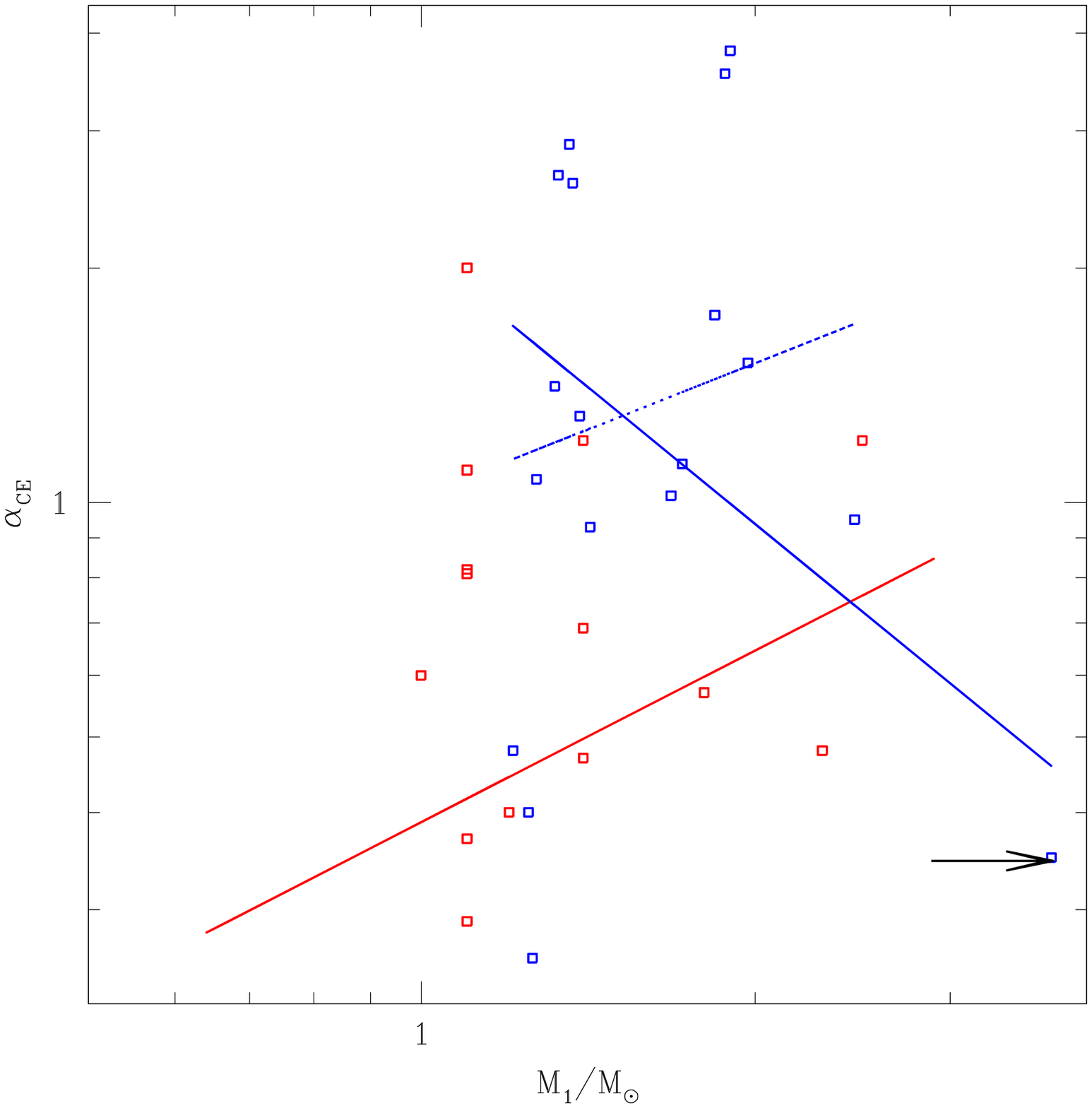}
    \includegraphics[scale=0.35]{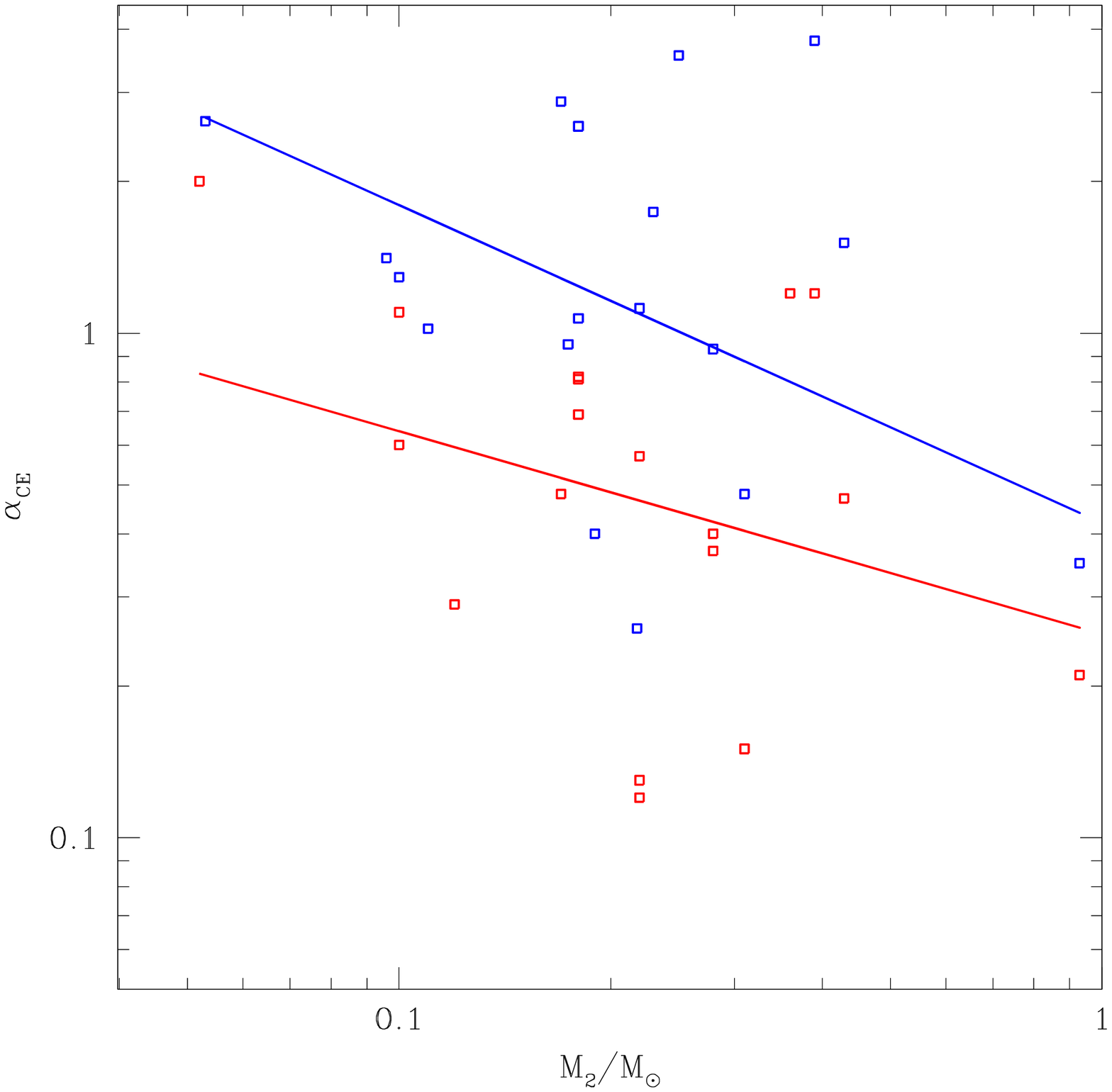}
   \includegraphics[scale=0.35]{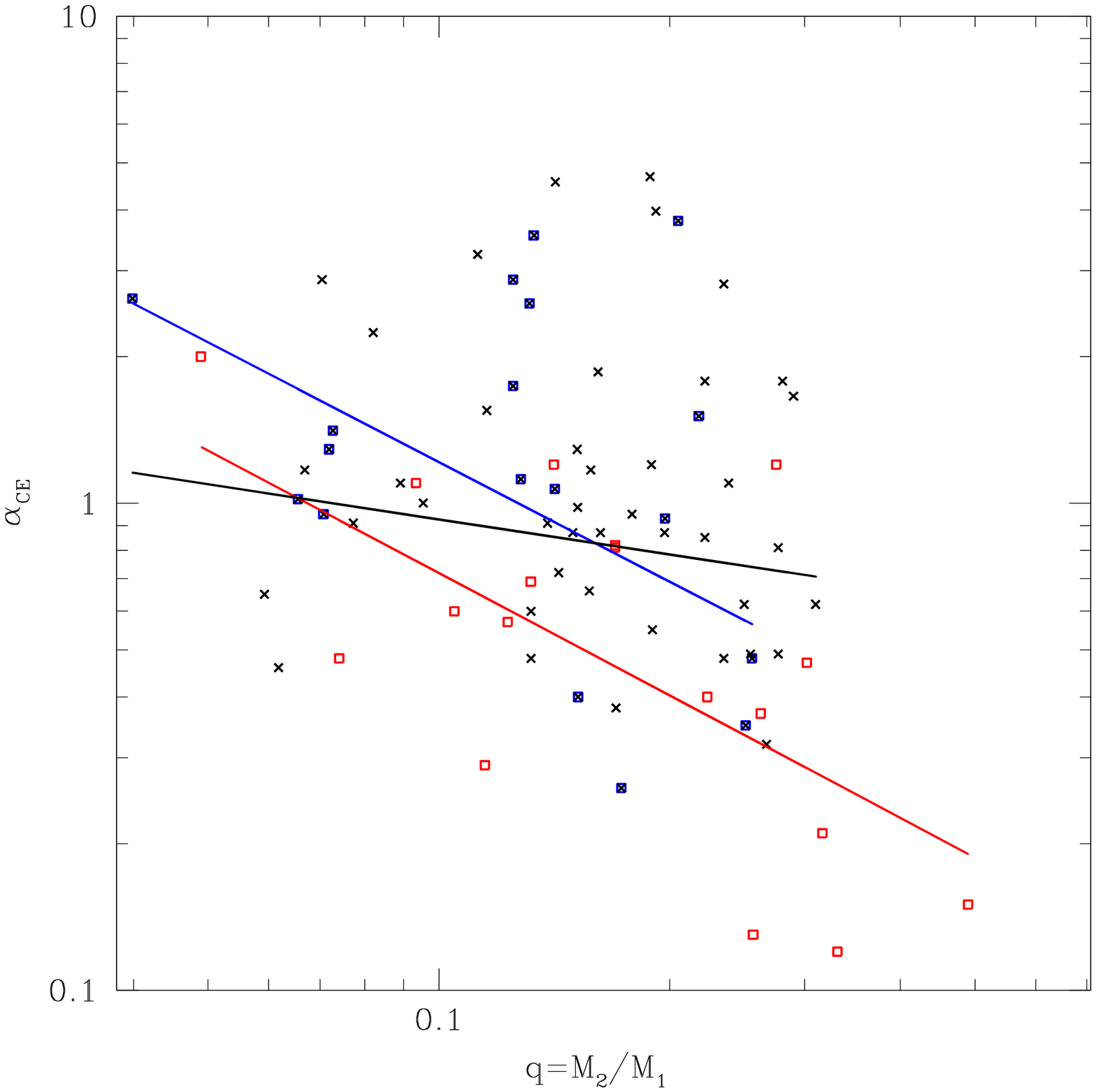}
    \caption{Reconstructed values of $\alpha_{\rmn{CE}}$ versus $M_{1}$
      (top left), $M_{2}$ (top right) and the mass ratio $q=M_{2}/M_{1}$
      (bottom middle). Blue data points correspond to our values, while red
      points have been calculated by \citet{DeMarco2011}. The black crosses
      in the bottom panel are our entire PCEB sample. The lines of best-fit
      are shown as the blue, red and black lines respectively. The dotted
      blue line in the top panel is a best fit to our data but with V471
      Tau (indicated by the arrow) omitted from the fit.}
    \label{ace_M1_M2_Q}
  \end{center}
\end{figure*}

\begin{table*}
  \begin{center}
  \begin{minipage}{140mm}
    \caption{Linear fit parameters for the data shown in the top-left,
      top-right and bottom panels of Fig. \ref{ace_M1_M2_Q}.}
    \begin{tabular}{@{}lccrcrcc@{}}
      \hline
      Study   &  $\epsilon_{0}$    &  $\epsilon_{1}$     &
      $\chi^{2}_{\rmn{linr}}$ & $\chi^{2}_{\rmn{cons}}$ &  $\mathcal{F}$  &
      $\mathscr{P}(\mathcal{F})$ & Reject
      $\mathscr{H}_{0}^{\mathcal{F}}$     \\
      \hline
      \multicolumn{8}{c}{$\alpha_{\rmn{CE}}$ versus $M_{1}$}\\
      \hline
      \citet{DeMarco2011}  &  -0.41$\pm$ 0.09 & 0.73$\pm$ 0.42 &
      16.28  &   19.32  &  3.0   & 0.1 &   No   \\
      This work (overlapping systems) &  0.32$\pm$0.15 & -1.16$\pm$0.40 & 34.07
      &  52.09    & 8.5   &  0.01   &   Yes  \\
      This work (no V471 Tau)  & 0.01$\pm$0.17  &  0.56$\pm$0.74  &  23.34
      &  24.24  &  0.6  &  0.5 & No   \\
      \hline
      \multicolumn{8}{c}{$\alpha_{\rmn{CE}}$ versus $M_{2}$}\\
      \hline
      \citet{DeMarco2011} & -0.59$\pm$0.22  &  -0.40$\pm$0.30 &  17.43  & 
      19.32  & 1.7    &  0.2  &  No     \\
      This work (overlapping systems)   &   -0.37$\pm$0.12  &
      -0.63$\pm$0.19  &  30.28  &  52.09    &  11.5 & 0.004  & Yes    \\
      \hline
      \multicolumn{8}{c}{$\alpha_{\rmn{CE}}$ versus $q$}\\
      \hline
      \citet{DeMarco2011}  &  -0.97$\pm$0.22  &  -0.84$\pm$0.27  &  12.00 & 19.32
       &  9.8 &  0.006  &  Yes\\       
      This work (overlapping systems) &  -0.73$\pm$0.27  &  -0.82$\pm$0.31
      & 36.48 & 52.09  &  6.8  &  0.02  & Yes    \\  
      This work (all systems)         &  -0.28$\pm$0.16  &  -0.24$\pm$0.17
      & 136.41 & 141.22 & 2.0  &  0.2   &   No   \\
      \hline
    \end{tabular}
    \label{compare_params}
  \end{minipage}
  \end{center}
\end{table*}

\subsection{Comparisons with \citet{DeMarco2011}}
\label{subsection:comparison_DM}

We now compare our values of $\alpha_{\rmn{CE}}$, calculated using
$\lambda=\lambda_{\rmn{g}}$, with those found by \citet{DeMarco2011} in
Fig. \ref{compare_DM}, for PCEBs that are common to both studies.

Fig. \ref{compare_DM} indicates that we obtain larger values of
$\alpha_{\rmn{CE}}$ by between a factor of 2 and 3. Reasons for this
disparity may be as follows. Firstly, \citet{DeMarco2011} use an
alternative expression for the envelope binding energy, given by the
left-hand side of their eqn. 6 which, for a given progenitor primary giant,
will give a smaller binding energy than that given by the left hand side of
the \citet{Webbink1984} formalism (c.f. our eqn. \ref{Ebind_4}). Hence, for
a given PCEB progenitor, a smaller value of $\alpha_{\rmn{CE}}$ will be
given if the \citet{DeMarco2011} formalism is used. Secondly, our values of
$\lambda_{\rmn{g}}$ are calculated from both the mass and radius of the
progenitor giant, while \citet{DeMarco2011} calculate their values purely
from the mass of the progenitor on the ZAMS. Hence, their method does not
take into account any variations of $\lambda$ due to the location of the
star (of a given mass) along the FGB or AGB, although they find that such
variations are small.

\citet{DeMarco2011} found that $\alpha_{\rmn{CE}}$ increases with
increasing $M_{1}$ but decreases with increasing $M_{2}$. Hence, they find
that $\alpha_{\rmn{CE}}$ decreases with increasing mass ratio,
$q=M_{2}/M_{1}$, from their PCEB sample. 

Fig. \ref{ace_M1_M2_Q} shows data points on the $\alpha_{\rmn{CE}}$-$M_{1}$
plane (top left panel), the $\alpha_{\rmn{CE}}$-$M_{2}$ plane (top right
panel) and on the $\alpha_{\rmn{CE}}$-$q$ plane (bottom middle panel). Blue
data points have been calculated in this present study, while red points
have been calculated by \citet{DeMarco2011}, again for the aforementioned
overlapping sample of PCEBs. The lines of best fit are shown as the blue
and red lines respectively.

The fit parameters to the data in each panel is summarised in Table
\ref{compare_params}, as are the (unreduced) chi-squared values,
$\chi^{2}_{\rmn{linr}}$ and $\chi^{2}_{\rmn{cons}}$, along with
$\mathcal{F}$ and $\mathscr{P}(\mathcal{F})$. We also indicate whether we
can reject $\mathscr{H}_{0}^{\mathcal{F}}$. The top-left panel of
Fig. \ref{ace_M1_M2_Q} shows that, in contrast to \citet{DeMarco2011}, a
linear fit to our data suggests a decreasing $\alpha_{\rmn{CE}}$ with
increasing $M_{1}$. However, the data point indicated by the arrow (which
corresponds to V471 Tau) may be overly influencing the fit. Indeed, the
standard deviation in $\alpha_{\rmn{CE}}$ for V471 Tau is the smallest in
the sub-sample. A linear fit to the PCEB sub-sample without V471 Tau is
given by the dotted blue line in the top-left panel of
Fig. \ref{ace_M1_M2_Q}. This is now consistent with the fit obtained by
\citet{DeMarco2011}. However, as Table \ref{compare_params} shows, we
cannot reject $\mathscr{H}_{0}^{\mathcal{F}}$.

On the other hand, the trends obtained from the linear fits to the blue
data points in the top-right and bottom panels are in line with those found
by \citet{DeMarco2011}. We see that we obtain a very good agreement with
\citet{DeMarco2011} for the value of $\epsilon_{1}$ in the
$\alpha_{\rmn{CE}}$-$q$ plane. Here, we can also reject
$\mathscr{H}_{0}^{\mathcal{F}}$ for both the red and blue data point on the
$\alpha_{\rmn{CE}}-q$ plane.

However, if we now perform a linear fit to our entire PCEB sample on the
$\alpha_{\rmn{CE}}-q$ plane (shown as the black crosses in the bottom panel
of Fig. \ref{ace_M1_M2_Q} and the black line respectively), then a trend
between $\alpha_{\rmn{CE}}$ and $q$ becomes less convincing. Indeed, we can
no longer reject $\mathscr{H}_{0}^{\mathcal{F}}$. It is therefore uncertain
whether the trend seen by \citet{DeMarco2011} is physical, or whether it is
due to the small PCEB sample which they used.

\section{Conclusions}

We reconstruct the common envelope (CE) phase for the current sample
of post-common envelope binaries (PCEBs) with observationally
determined component masses and orbital periods. We perform
multi-regression analysis on our reconstructed values of the CE
ejection efficiency, $\alpha_{\rmn{CE}}$, in order to search for
correlations between $\alpha_{\rmn{CE}}$ and the binary parameters
(both the progenitor and observed parameters). This analysis is
carried out with and without consideration for the internal energy of
the progenitor giants' envelope.

If the internal energy of the progenitor giants' envelope is taken into
account, a statistically significant correlation (in terms of the
progenitor parameters) is found between $\alpha_{\rmn{CE}}$ and the white
dwarf mass, $M_{\rmn{WD}}$, only. We find that $\alpha_{\rmn{CE}}$
decreases with increasing $M_{\rmn{WD}}$.

If we do not consider the envelopes' internal energy, we find the most
statistically significant correlation exists between $\alpha_{\rmn{CE}}$
and the orbital period at the start of the CE phase,
$P_{\rmn{CE,i}}$. Specifically, we find that $\alpha_{\rmn{CE}}$ decreases
with increasing $P_{\rmn{CE,i}}$. In terms of the PCEB parameters, whether
we consider the internal energy or not, we find a significant correlation
between $\alpha_{\rmn{CE}}$ and $M_{\rmn{WD}}$, the orbital period
immediately after the CE phase, $P_{\rmn{CE}}$ and the secondary mass,
$M_2$. While our values of $\alpha_{\rmn{CE}}$, and the aforementioned
correlations, cannot make definite physical claims regarding the CE phase,
they will be nonetheless useful to constrain the micro-physics modelled in
detailed CE simulations.

We re-iterate that uncertainties in the measured white dwarf masses result
in uncertainties in the progenitor binary parameters, and in turn the
values of $\alpha_{\rmn{CE}}$. Future investigations into reconstructing
the CE phase would therefore greatly benefit from more accurate
determinations of PCEB white dwarf masses, for example from eclipsing
systems.

If the internal energy of the progenitor primary envelope is
considered, this brings the values of $\alpha_{\rmn{CE}}$ more in line
with the canonical range of $0<\alpha_{\rmn{CE}}\leq{1}$. However, we
still require $\alpha_{\rmn{CE}} \ga 1$ to account for PCEBs with
brown dwarf secondaries, i.e. with $M_{2}\la{0.1}$ M$_{\odot}$.

Finally, a large majority of possible progenitor primaries of IK
Pegasi will have envelopes that have a positive binding energy. We
suggest that IK Peg may have avoided the CE phase, and instead mass
was lost from the system via radiation pressure driven winds.

\section*{acknowledgements}

PJD acknowledges financial support from the Communaut\'{e} fran\c{c}aise de
Belgique - Actions de recherche Concert\'{e}es, and from the Institut
d'Astronomie et d'Astrophysique at the Universit\'{e} Libre de Bruxelles
(ULB). We would also like to thank the anonymous referee for the positive
feedback.

%
%

\begin{thebibliography}{}

\bibitem[\protect\citeauthoryear{{Aungwerojwit}, {G{\"a}nsicke},
  {Rodr{\'{\i}}guez-Gil}, {Hagen}, {Giannakis}, {Papadimitriou}, {Allende
  Prieto} \& {Engels}}{{Aungwerojwit} et~al.}{2007}]{aungwerojwit07}
{Aungwerojwit} A.,  {G{\"a}nsicke} B.~T.,  {Rodr{\'{\i}}guez-Gil} P.,  {Hagen}
  H.-J.,  {Giannakis} O.,  {Papadimitriou} C.,  {Allende Prieto} C.,
  {Engels} D.,  2007, A\&A, 469, 297

\bibitem[\protect\citeauthoryear{{Beer}, {Dray}, {King} \& {Wynn}}{{Beer}
  et~al.}{2007}]{Beer2007}
{Beer} M.~E.,  {Dray} L.~M.,  {King} A.~R.,    {Wynn} G.~A.,  2007, MNRAS, 375,
  1000

\bibitem[\protect\citeauthoryear{{Bleach}, {Wood}, {Smalley} \&
  {Catal{\'a}n}}{{Bleach} et~al.}{2002}]{Bleach2002}
{Bleach} J.~N.,  {Wood} J.~H.,  {Smalley} B.,    {Catal{\'a}n} M.~S.,  2002,
  MNRAS, 336, 611

\bibitem[\protect\citeauthoryear{{Bruch} \& {Diaz}}{{Bruch} \&
  {Diaz}}{1999}]{Bruch1999}
{Bruch} A.,  {Diaz} M.~P.,  1999, A\&A, 351, 573

\bibitem[\protect\citeauthoryear{{Catalan}, {Davey}, {Sarna}, {Connon-Smith} \&
  {Wood}}{{Catalan} et~al.}{1994}]{catalan94}
{Catalan} M.~S.,  {Davey} S.~C.,  {Sarna} M.~J.,  {Connon-Smith} R.,    {Wood}
  J.~H.,  1994, MNRAS, 269, 879

\bibitem[\protect\citeauthoryear{{Davis}, {Kolb} \& {Willems}}{{Davis}
  et~al.}{2010}]{Davis2010}
{Davis} P.~J.,  {Kolb} U.,    {Willems} B.,  2010, MNRAS, 403, 179

\bibitem[\protect\citeauthoryear{{Davis}, {Kolb}, {Willems} \&
  {G{\"a}nsicke}}{{Davis} et~al.}{2008}]{Davis2008}
{Davis} P.~J.,  {Kolb} U.,  {Willems} B.,    {G{\"a}nsicke} B.~T.,  2008,
  MNRAS, 389, 1563

\bibitem[\protect\citeauthoryear{{de Kool}}{{de Kool}}{1992}]{deKool1992}
{de Kool} M.,  1992, {A\&{A}}, 261, 188

\bibitem[\protect\citeauthoryear{{de Kool} \& {Ritter}}{{de Kool} \&
  {Ritter}}{1993}]{deKool1993}
{de Kool} M.,  {Ritter} H.,  1993, {A\&A}, 267, 397

\bibitem[\protect\citeauthoryear{{De Marco}, {Passy}, {Moe}, {Herwig}, {Mac
  Low} \& {Paxton}}{{De Marco} et~al.}{2011}]{DeMarco2011}
{De Marco} O.,  {Passy} J.,  {Moe} M.,  {Herwig} F.,  {Mac Low} M.,    {Paxton}
  B.,  2011, MNRAS, 411, 2277

\bibitem[\protect\citeauthoryear{{Dermine}, {Jorissen}, {Siess} \&
  {Frankowski}}{{Dermine} et~al.}{2009}]{Dermine2009}
{Dermine} T.,  {Jorissen} A.,  {Siess} L.,    {Frankowski} A.,  2009, A\&A,
  507, 891

\bibitem[\protect\citeauthoryear{{Dewi} \& {Tauris}}{{Dewi} \&
  {Tauris}}{2000}]{Dewi2000}
{Dewi} J.~D.~M.,  {Tauris} T.~M.,  2000, AAP, 360, 1043

\bibitem[\protect\citeauthoryear{{Fulbright}, {Liebert}, {Bergeron} \&
  {Green}}{{Fulbright} et~al.}{1993}]{fulbright93}
{Fulbright} M.~S.,  {Liebert} J.,  {Bergeron} P.,    {Green} R.,  1993, ApJ,
  406, 240

\bibitem[\protect\citeauthoryear{{Good}, {Barstow}, {Burleigh}, {Dobbie} \&
  {Holberg}}{{Good} et~al.}{2005}]{good05}
{Good} S.~A.,  {Barstow} M.~A.,  {Burleigh} M.~R.,  {Dobbie} P.~D.,
  {Holberg} J.~B.,  2005, MNRAS, 364, 1082

\bibitem[\protect\citeauthoryear{{Han}, {Podsiadlowski}, {Maxted}, {Marsh} \&
  {Ivanova}}{{Han} et~al.}{2002}]{Han2002}
{Han} Z.,  {Podsiadlowski} P.,  {Maxted} P.~F.~L.,  {Marsh} T.~R.,    {Ivanova}
  N.,  2002, MNRAS, 336, 449

\bibitem[\protect\citeauthoryear{{Harpaz}}{{Harpaz}}{1998}]{Harpaz1998}
{Harpaz} A.,  1998, ApJ, 498, 293

\bibitem[\protect\citeauthoryear{{Hjellming} \& {Webbink}}{{Hjellming} \&
  {Webbink}}{1987}]{Hjellming1987}
{Hjellming} M.~S.,  {Webbink} R.~F.,  1987, ApJ, 318, 794

\bibitem[\protect\citeauthoryear{{Hurley}, {Pols} \& {Tout}}{{Hurley}
  et~al.}{2000}]{Hurley2000}
{Hurley} J.~R.,  {Pols} O.~R.,    {Tout} C.~A.,  2000, MNRAS, 315, 543

\bibitem[\protect\citeauthoryear{{Hurley}, {Tout} \& {Pols}}{{Hurley}
  et~al.}{2002}]{Hurley2002}
{Hurley} J.~R.,  {Tout} C.~A.,    {Pols} O.~R.,  2002, MNRAS, 329, 897

\bibitem[\protect\citeauthoryear{{Iben} Jr. \& {Tutukov}}{{Iben} \&
  {Tutukov}}{1984}]{Iben1984}
{Iben} Jr. I.,  {Tutukov} A.~V.,  1984, ApJ, 284, 719

\bibitem[\protect\citeauthoryear{{Iben} \& {Livio}}{{Iben} \&
  {Livio}}{1993}]{Iben1993}
{Iben} I.~J.,  {Livio} M.,  1993, PASP, 105, 1373

\bibitem[\protect\citeauthoryear{{Ivanova}}{{Ivanova}}{2011}]{Ivanova2011}
{Ivanova} N.,  2011, APJ, 730, 76

\bibitem[\protect\citeauthoryear{{Ivanova} \& {Chaichenets}}{{Ivanova} \&
  {Chaichenets}}{2011}]{Ivanova2011b}
{Ivanova} N.,  {Chaichenets} S.,  2011, APJL, 731, L36+

\bibitem[\protect\citeauthoryear{{Kami{\'n}ski}, {Ruci{\'n}ski}, {Matthews},
  {Kuschnig}, {Rowe}, {Guenther}, {Moffat}, {Sasselov}, {Walker} \&
  {Weiss}}{{Kami{\'n}ski} et~al.}{2007}]{Kaminsky2007}
{Kami{\'n}ski} K.~Z.,  {Ruci{\'n}ski} S.~M.,  {Matthews} J.~M.,  {Kuschnig} R.,
   {Rowe} J.~F.,  {Guenther} D.~B.,  {Moffat} A.~F.~J.,  {Sasselov} D.,
  {Walker} G.~A.~H.,    {Weiss} W.~W.,  2007, AJ, 134, 1206

\bibitem[\protect\citeauthoryear{{Kawka} \& {Vennes}}{{Kawka} \&
  {Vennes}}{2003}]{Kawka2003}
{Kawka} A.,  {Vennes} S.,  2003, AJ, 125, 1444

\bibitem[\protect\citeauthoryear{{Kawka} \& {Vennes}}{{Kawka} \&
  {Vennes}}{2005}]{Kawka2005}
{Kawka} A.,  {Vennes} S.,  2005, Ap\&SS, 296, 481

\bibitem[\protect\citeauthoryear{{Kawka}, {Vennes}, {Dupuis}, {Chayer} \&
  {Lanz}}{{Kawka} et~al.}{2008}]{kawka08}
{Kawka} A.,  {Vennes} S.,  {Dupuis} J.,  {Chayer} P.,    {Lanz} T.,  2008, ApJ,
  675, 1518

\bibitem[\protect\citeauthoryear{{Kawka}, {Vennes}, {Koch} \&
  {Williams}}{{Kawka} et~al.}{2002}]{kawka02}
{Kawka} A.,  {Vennes} S.,  {Koch} R.,    {Williams} A.,  2002, AJ, 124, 2853

\bibitem[\protect\citeauthoryear{{Kolb} \& {Baraffe}}{{Kolb} \&
  {Baraffe}}{1999}]{Kolb1999}
{Kolb} U.,  {Baraffe} I.,  1999, MNRAS, 309, 1034

\bibitem[\protect\citeauthoryear{{Kroupa}, {Tout} \& {Gilmore}}{{Kroupa}
  et~al.}{1993}]{Kroupa1993}
{Kroupa} P.,  {Tout} C.~A.,    {Gilmore} G.,  1993, MNRAS, 262, 545

\bibitem[\protect\citeauthoryear{{Kudritzki} \& {Reimers}}{{Kudritzki} \&
  {Reimers}}{1978}]{Kudritzki1978}
{Kudritzki} R.~P.,  {Reimers} D.,  1978, A\&A, 70, 227

\bibitem[\protect\citeauthoryear{{Landsman}, {Simon} \& {Bergeron}}{{Landsman}
  et~al.}{1993}]{lsb93}
{Landsman} W.,  {Simon} T.,    {Bergeron} P.,  1993, PASP, 105, 841

\bibitem[\protect\citeauthoryear{{Lanning} \& {Pesch}}{{Lanning} \&
  {Pesch}}{1981}]{lp81}
{Lanning} H.~H.,  {Pesch} P.,  1981, ApJ, 244, 280

\bibitem[\protect\citeauthoryear{{Maxted}, {Marsh}, {Heber}, {Morales-Rueda},
  {North} \& {Lawson}}{{Maxted} et~al.}{2002}]{Maxted2002a}
{Maxted} P.~F.~L.,  {Marsh} T.~R.,  {Heber} U.,  {Morales-Rueda} L.,  {North}
  R.~C.,    {Lawson} W.~A.,  2002, MNRAS, 333, 231

\bibitem[\protect\citeauthoryear{{Maxted}, {Marsh}, {Morales-Rueda}, {Barstow},
  {Dobbie}, {Schreiber}, {Dhillon} \& {Brinkworth}}{{Maxted}
  et~al.}{2004}]{maxted04}
{Maxted} P.~F.~L.,  {Marsh} T.~R.,  {Morales-Rueda} L.,  {Barstow} M.~A.,
  {Dobbie} P.~D.,  {Schreiber} M.~R.,  {Dhillon} V.~S.,    {Brinkworth} C.~S.,
  2004, MNRAS, 355, 1143

\bibitem[\protect\citeauthoryear{{Maxted}, {Marsh}, {Moran}, {Dhillon} \&
  {Hilditch}}{{Maxted} et~al.}{1998}]{maxted98}
{Maxted} P.~F.~L.,  {Marsh} T.~R.,  {Moran} C.,  {Dhillon} V.~S.,    {Hilditch}
  R.~W.,  1998, MNRAS, 300, 1225

\bibitem[\protect\citeauthoryear{{Maxted}, {Napiwotzki}, {Dobbie} \&
  {Burleigh}}{{Maxted} et~al.}{2006}]{maxted06}
{Maxted} P.~F.~L.,  {Napiwotzki} R.,  {Dobbie} P.~D.,    {Burleigh} M.~R.,
  2006, Nature, 442, 543

\bibitem[\protect\citeauthoryear{{Maxted}, {Napiwotzki}, {Marsh}, {Burleigh},
  {Dobbie}, {Hogan} \& {Nelemans}}{{Maxted} et~al.}{2007a}]{maxted07}
{Maxted} P.~F.~L.,  {Napiwotzki} R.,  {Marsh} T.~R.,  {Burleigh} M.~R.,
  {Dobbie} P.~D.,  {Hogan} E.,    {Nelemans} G.,  {2007a}, in {Napiwotzki} R.,
  {Burleigh} M.~R.,  eds, 15th European Workshop on White Dwarfs Vol.~372 of
  Astronomical Society of the Pacific Conference Series, {Follow-up
  Observations of SPY White Dwarf + M-Dwarf Binaries}.
pp 471--+

\bibitem[\protect\citeauthoryear{{Maxted}, {O'Donoghue}, {Morales-Rueda},
  {Napiwotzki} \& {Smalley}}{{Maxted} et~al.}{2007b}]{Maxted2007}
{Maxted} P.~F.~L.,  {O'Donoghue} D.,  {Morales-Rueda} L.,  {Napiwotzki} R.,
  {Smalley} B.,  {2007b}, MNRAS, 376, 919

\bibitem[\protect\citeauthoryear{{Morales-Rueda}, {Marsh}, {Maxted},
  {Nelemans}, {Karl}, {Napiwotzki} \& {Moran}}{{Morales-Rueda}
  et~al.}{2005}]{morales05}
{Morales-Rueda} L.,  {Marsh} T.~R.,  {Maxted} P.~F.~L.,  {Nelemans} G.,  {Karl}
  C.,  {Napiwotzki} R.,    {Moran} C.~K.~J.,  2005, MNRAS, 359, 648

\bibitem[\protect\citeauthoryear{{Nebot G{\'o}mez-Mor{\'a}n}, {Schwope},
  {Schreiber} \& {G{\"a}nsicke}}{{Nebot G{\'o}mez-Mor{\'a}n}
  et~al.}{2009}]{Moran2009}
{Nebot G{\'o}mez-Mor{\'a}n} A.,  {Schwope} A.~D.,  {Schreiber} M.~R.,
  {G{\"a}nsicke} B.~T.,  2009, Journal of Physics Conference Series, 172,
  012027

\bibitem[\protect\citeauthoryear{{Nelemans} \& {Tout}}{{Nelemans} \&
  {Tout}}{2005}]{Nelemans2005}
{Nelemans} G.,  {Tout} C.~A.,  2005, MNRAS, 356, 753

\bibitem[\protect\citeauthoryear{{Nelemans}, {Verbunt}, {Yungelson} \&
  {Portegies Zwart}}{{Nelemans} et~al.}{2000}]{Nelemans2000}
{Nelemans} G.,  {Verbunt} F.,  {Yungelson} L.~R.,    {Portegies Zwart} S.~F.,
  2000, A\&A, 360, 1011

\bibitem[\protect\citeauthoryear{{O'Brien}, {Bond} \& {Sion}}{{O'Brien}
  et~al.}{2001}]{obrien01}
{O'Brien} M.~S.,  {Bond} H.~E.,    {Sion} E.~M.,  2001, ApJ, 563, 971

\bibitem[\protect\citeauthoryear{{O'Donoghue}, {Koen}, {Kilkenny}, {Stobie},
  {Koester}, {Bessell}, {Hambly} \& {MacGillivray}}{{O'Donoghue}
  et~al.}{2003}]{odonoghue03}
{O'Donoghue} D.,  {Koen} C.,  {Kilkenny} D.,  {Stobie} R.~S.,  {Koester} D.,
  {Bessell} M.~S.,  {Hambly} N.,    {MacGillivray} H.,  2003, MNRAS, 345, 506

\bibitem[\protect\citeauthoryear{{Paczynski}}{{Paczynski}}{1976}]{Paczynski197%
6}
{Paczynski} B.,  1976, in {P.~Eggleton, S.~Mitton, \& J.~Whelan} ed., Structure
  and Evolution of Close Binary Systems Vol.~73 of IAU Symposium, {Common
  Envelope Binaries}.
{Reidel Publishing Co., Dordrecht}, pp 75--+

\bibitem[\protect\citeauthoryear{{Parsons}, {Marsh}, {Copperwheat}, {Dhillon},
  {Littlefair}, {Hickman}, {Maxted}, {G{\"a}nsicke}, {Unda-Sanzana}, {Colque},
  {Barraza}, {S{\'a}nchez} \& {Monard}}{{Parsons} et~al.}{2010}]{Parsons2010}
{Parsons} S.~G.,  {Marsh} T.~R.,  {Copperwheat} C.~M.,  {Dhillon} V.~S.,
  {Littlefair} S.~P.,  {Hickman} R.~D.~G.,  {Maxted} P.~F.~L.,  {G{\"a}nsicke}
  B.~T.,  {Unda-Sanzana} E.,  {Colque} J.~P.,  {Barraza} N.,  {S{\'a}nchez} N.,
     {Monard} L.~A.~G.,  2010, MNRAS, 407, 2362

\bibitem[\protect\citeauthoryear{{Paxton}}{{Paxton}}{2004}]{Paxton2004}
{Paxton} B.,  2004, PASP, 116, 699

\bibitem[\protect\citeauthoryear{{Politano}}{{Politano}}{2004}]{Politano2004}
{Politano} M.,  2004, ApJ, 604, 817

\bibitem[\protect\citeauthoryear{{Politano} \& {Weiler}}{{Politano} \&
  {Weiler}}{2007}]{Politano2007}
{Politano} M.,  {Weiler} K.~P.,  2007, ApJ, 665, 663

\bibitem[\protect\citeauthoryear{{Pols}, {Schroder}, {Hurley}, {Tout} \&
  {Eggleton}}{{Pols} et~al.}{1998}]{Pols1998}
{Pols} O.~R.,  {Schroder} K.-P.,  {Hurley} J.~R.,  {Tout} C.~A.,    {Eggleton}
  P.~P.,  1998, MNRAS, 298, 525

\bibitem[\protect\citeauthoryear{{Pyrzas}, {G{\"a}nsicke}, {Marsh},
  {Aungwerojwit}, {Rebassa-Mansergas}, {Rodr{\'{\i}}guez-Gil}, {Southworth},
  {Schreiber}, {Nebot Gomez-Moran} \& {Koester}}{{Pyrzas}
  et~al.}{2009}]{Pyrzas2009}
{Pyrzas} S.,  {G{\"a}nsicke} B.~T.,  {Marsh} T.~R.,  {Aungwerojwit} A.,
  {Rebassa-Mansergas} A.,  {Rodr{\'{\i}}guez-Gil} P.,  {Southworth} J.,
  {Schreiber} M.~R.,  {Nebot Gomez-Moran} A.,    {Koester} D.,  2009, MNRAS,
  394, 978

\bibitem[\protect\citeauthoryear{{Raguzova}, {Shugarov} \&
  {Ketsaris}}{{Raguzova} et~al.}{2003}]{Raguzova2003}
{Raguzova} N.~V.,  {Shugarov} S.~Y.,    {Ketsaris} N.~A.,  2003, Astronomy
  Reports, 47, 492

\bibitem[\protect\citeauthoryear{{Rappaport}, {Verbunt} \& {Joss}}{{Rappaport}
  et~al.}{1983}]{Rappaport1983}
{Rappaport} S.,  {Verbunt} F.,    {Joss} P.~C.,  1983, {ApJ}, 275, 713

\bibitem[\protect\citeauthoryear{{Rebassa-Mansergas}, {G{\"a}nsicke},
  {Schreiber}, {Southworth} \& {Schwope}}{{Rebassa-Mansergas}
  et~al.}{2008}]{Mansergas2008}
{Rebassa-Mansergas} A.,  {G{\"a}nsicke} B.~T.,  {Schreiber} M.~R.,
  {Southworth} J.,    {Schwope} A.~D.,  2008, MNRAS, 390, 1635

\bibitem[\protect\citeauthoryear{{Ritter} \& {Kolb}}{{Ritter} \&
  {Kolb}}{2003}]{Ritter2003}
{Ritter} H.,  {Kolb} U.,  2003, A\&A, 404, 301

\bibitem[\protect\citeauthoryear{{Saffer}, {Wade}, {Liebert}, {Green}, {Sion},
  {Bechtold}, {Foss} \& {Kidder}}{{Saffer} et~al.}{1993}]{saffer93}
{Saffer} R.~A.,  {Wade} R.~A.,  {Liebert} J.,  {Green} R.~F.,  {Sion} E.~M.,
  {Bechtold} J.,  {Foss} D.,    {Kidder} K.,  1993, ApJ, 105, 1945

\bibitem[\protect\citeauthoryear{{Sandquist}, {Taam} \& {Burkert}}{{Sandquist}
  et~al.}{2000}]{Sandquist2000}
{Sandquist} E.~L.,  {Taam} R.~E.,    {Burkert} A.,  2000, ApJ, 533, 984

\bibitem[\protect\citeauthoryear{{Schreiber} \& {G{\"a}nsicke}}{{Schreiber} \&
  {G{\"a}nsicke}}{2003}]{Schreiber2003}
{Schreiber} M.~R.,  {G{\"a}nsicke} B.~T.,  2003, A\&A, 406, 305

\bibitem[\protect\citeauthoryear{{Schreiber}, {G{\"a}nsicke}, {Southworth},
  {Schwope} \& {Koester}}{{Schreiber} et~al.}{2008}]{Schreiber2008}
{Schreiber} M.~R.,  {G{\"a}nsicke} B.~T.,  {Southworth} J.,  {Schwope} A.~D.,
   {Koester} D.,  2008, A\&A, 484, 441

\bibitem[\protect\citeauthoryear{{Shimanskii} \& {Borisov}}{{Shimanskii} \&
  {Borisov}}{2002}]{Shimanskii2002}
{Shimanskii} V.~V.,  {Borisov} N.~V.,  2002, Astronomy Reports, 46, 406

\bibitem[\protect\citeauthoryear{{Shimanskii}, {Borisov}, {Pozdnyakova},
  {Bikmaev}, {Vlasyuk}, {Sakhibullin} \& {Spiridonova}}{{Shimanskii}
  et~al.}{2008}]{Shimanskii2008}
{Shimanskii} V.~V.,  {Borisov} N.~V.,  {Pozdnyakova} S.~A.,  {Bikmaev} I.~F.,
  {Vlasyuk} V.~V.,  {Sakhibullin} N.~A.,    {Spiridonova} O.~I.,  2008,
  Astronomy Reports, 52, 558

\bibitem[\protect\citeauthoryear{{Shimansky}, {Borisov} \&
  {Shimanskaya}}{{Shimansky} et~al.}{2003}]{shimansky03}
{Shimansky} V.~V.,  {Borisov} N.~V.,    {Shimanskaya} N.~N.,  2003, Astronomy
  Reports, 47, 763

\bibitem[\protect\citeauthoryear{{Shimansky}, {Pozdnyakova}, {Borisov},
  {Bikmaev}, {Vlasyuk}, {Spiridonova}, {Galeev} \& {Mel'Nikov}}{{Shimansky}
  et~al.}{2009}]{Shimansky2009}
{Shimansky} V.~V.,  {Pozdnyakova} S.~A.,  {Borisov} N.~V.,  {Bikmaev} I.~F.,
  {Vlasyuk} V.~V.,  {Spiridonova} O.~I.,  {Galeev} A.~I.,    {Mel'Nikov} S.~S.,
   2009, Astrophysical Bulletin, 64, 349

\bibitem[\protect\citeauthoryear{{Somers}, {Lockley}, {Naylor} \&
  {Wood}}{{Somers} et~al.}{1996}]{Somers1996}
{Somers} M.~W.,  {Lockley} J.~J.,  {Naylor} T.,    {Wood} J.~H.,  1996, MNRAS,
  280, 1277

\bibitem[\protect\citeauthoryear{{Spruit} \& {Ritter}}{{Spruit} \&
  {Ritter}}{1983}]{Spruit1983}
{Spruit} H.~C.,  {Ritter} H.,  1983, {Astronomy and Astrophysics}, 124, 267

\bibitem[\protect\citeauthoryear{{Stauffer}}{{Stauffer}}{1987}]{stauffer87}
{Stauffer} J.~R.,  1987, AJ, 94, 996

\bibitem[\protect\citeauthoryear{{Tappert}, {G{\"a}nsicke}, {Schmidtobreick},
  {Aungwerojwit}, {Mennickent} \& {Koester}}{{Tappert}
  et~al.}{2007}]{tappert07}
{Tappert} C.,  {G{\"a}nsicke} B.~T.,  {Schmidtobreick} L.,  {Aungwerojwit} A.,
  {Mennickent} R.~E.,    {Koester} D.,  2007, A\&A, 474, 205

\bibitem[\protect\citeauthoryear{{Terman} \& {Taam}}{{Terman} \&
  {Taam}}{1996}]{Terman1996}
{Terman} J.~L.,  {Taam} R.~E.,  1996, ApJ, 458, 692

\bibitem[\protect\citeauthoryear{{van den Besselaar}, {Greimel},
  {Morales-Rueda}, {Nelemans}, {Thorstensen}, {Marsh}, {Dhillon}, {Robb},
  {Balam}, {Guenther}, {Kemp}, {Augusteijn} \& {Groot}}{{van den Besselaar}
  et~al.}{2007}]{besselaar07}
{van den Besselaar} E.~J.~M.,  {Greimel} R.,  {Morales-Rueda} L.,  {Nelemans}
  G.,  {Thorstensen} J.~R.,  {Marsh} T.~R.,  {Dhillon} V.~S.,  {Robb} R.~M.,
  {Balam} D.~D.,  {Guenther} E.~W.,  {Kemp} J.,  {Augusteijn} T.,    {Groot}
  P.~J.,  2007, A\&A, 466, 1031

\bibitem[\protect\citeauthoryear{{Vassiliadis} \& {Wood}}{{Vassiliadis} \&
  {Wood}}{1993}]{Vassiliadis1993}
{Vassiliadis} E.,  {Wood} P.~R.,  1993, ApJ, 413, 641

\bibitem[\protect\citeauthoryear{{Vennes}, {Christian} \&
  {Thorstensen}}{{Vennes} et~al.}{1998}]{Vennes1998}
{Vennes} S.,  {Christian} D.~J.,    {Thorstensen} J.~R.,  1998, ApJ, 502, 763

\bibitem[\protect\citeauthoryear{{Vennes} \& {Thorstensen}}{{Vennes} \&
  {Thorstensen}}{1994}]{vt94}
{Vennes} S.,  {Thorstensen} J.~R.,  1994, AJ, 108, 1881

\bibitem[\protect\citeauthoryear{{Vennes}, {Thorstensen} \&
  {Polomski}}{{Vennes} et~al.}{1999}]{vennes99}
{Vennes} S.,  {Thorstensen} J.~R.,    {Polomski} E.~F.,  1999, ApJ, 523, 386

\bibitem[\protect\citeauthoryear{{Webbink}}{{Webbink}}{1984}]{Webbink1984}
{Webbink} R.~F.,  1984, APJ, 277, 355

\bibitem[\protect\citeauthoryear{{Webbink}}{{Webbink}}{2008}]{Webbink2008}
{Webbink} R.~F.,  2008, in {E.~F.~Milone, D.~A.~Leahy, \& D.~W.~Hobill} ed.,
  Astrophysics and Space Science Library Vol.~352 of Astrophysics and Space
  Science Library, {Common Envelope Evolution Redux}.
{Springer, Berlin}, pp 233--+

\bibitem[\protect\citeauthoryear{{Willems} \& {Kolb}}{{Willems} \&
  {Kolb}}{2002}]{Willems2002}
{Willems} B.,  {Kolb} U.,  2002, MNRAS, 337, 1004

\bibitem[\protect\citeauthoryear{{Willems} \& {Kolb}}{{Willems} \&
  {Kolb}}{2004}]{Willems2004}
{Willems} B.,  {Kolb} U.,  2004, AAP, 419, 1057

\bibitem[\protect\citeauthoryear{{Yorke}, {Bodenheimer} \& {Taam}}{{Yorke}
  et~al.}{1995}]{Yorke95}
{Yorke} H.~W.,  {Bodenheimer} P.,    {Taam} R.~E.,  1995, ApJ, 451, 308

\bibitem[\protect\citeauthoryear{{Zorotovic}, {Schreiber}, {G{\"a}nsicke} \&
  {Nebot G{\'o}mez-Mor{\'a}n}}{{Zorotovic} et~al.}{2010}]{Zorotovic2010}
{Zorotovic} M.,  {Schreiber} M.~R.,  {G{\"a}nsicke} B.~T.,    {Nebot
  G{\'o}mez-Mor{\'a}n} A.,  2010, A\&A, 520, A86+

\end{thebibliography}

\appendix

\section{Initial distribution functions}

We use three standard distribution functions to calculate the
formation probability of the zero-age main sequence binary system.
The initial mass function (IMF) is given by \citep{Kroupa1993}
\begin{equation}f(M_{1,\rmn{i}}) = \left\{
\begin{array}{l l}
  0 & \quad \mbox{$M_{1,\rmn{i}}/\rmn{M_{\odot}}<0.1,$}\\ 0.29056M_{1}^{-1.3}
  & \quad \mbox{$0.1\leq{M_{1,\rmn{i}}/\rmn{M_{\odot}}}<0.5,$}\\
  0.15571M_{1}^{-2.2} & \quad
  \mbox{$0.5\leq{M_{1,\rmn{i}}/\rmn{M_{\odot}}}<1.0,$}\\ 0.15571M_{1}^{-2.7} &
  \quad \mbox{$1.0\leq{M_{1,\rmn{i}}/\rmn{M_{\odot}}},$}\\ \end{array}
  \right.
\label{IMF}
\end{equation}
The initial orbital separation distribution (IOSD) function,
$h(a_{\rmn{i}})$, is given by \citep{Iben1984,Hurley2002}
\begin{equation}h(a_{\rmn{i}}) = \left\{
\begin{array}{l l}
  0 & \quad
  \mbox{$a_{\rmn{i}}/\rmn{R_{\odot}}<3\:\:\rmn{or}\:\:a_{\rmn{i}}/\rmn{R_{\odot}}>10^{6},$}\\
  0.078636a_{\rmn{i}}^{-1} & \quad
  \mbox{$3\leq{a_{\rmn{i}}/\rmn{R_{\odot}}}\leq{10^{6}}.$}\\
  \end{array} \right.
\label{IOSD}\
\end{equation}
Finally, the initial mass ratio distribution function,
$g(q_{\rmn{i}})$, can be calculated from
\begin{equation}
q(q_{\rmn{i}})=\mu q_{\rmn{i}}^{\nu},
\label{IMRD}
\end{equation}
where $\mu$ is a normalisation factor, and $\nu$ is a free parameter.

\section{The F-Test}
\label{subsection:F_test}

The F-test determines how much an additional term has improved the value of
the reduced chi-squared value. If the chi-squared values from fits using
$m$ and $m+1$ parameters are $\chi^2_{m}$ and $\chi^{2}_{m+1}$
respectively, then the F-statistic, $\mathcal{F}$, is determined using
\begin{equation}
\mathcal{F}=\frac{\chi^{2}_{m}-\chi^{2}_{m+1}}{\chi^{2}_{m+1}/(N-m-1)},
\label{F-value}
\end{equation}
where $N$ is the number of data points. We then calculate the probability,
$\mathscr{P}(\mathcal{F})$, of exceeding the calculated value of
$\mathcal{F}$.

We can reject the null hypothesis that the additional term is warranted,
$\mathscr{H}^{\mathcal{F}}_{0}$, if the value of $\mathscr{P}(\mathcal{F})$
is less than some significance level, $\alpha$, i.e.
$\mathscr{P}(\mathcal{F})<\alpha$.

\section{Observed and derived parameters for the observed sample of PCEBs}

\begin{table*}
  \centering
  \begin{minipage}{170mm}
    \caption{A table showing the orbital parameters of the observed
      sample of PCEBs from the SDSS. Also given are the orbital
      periods immediately after the CE phase, $P_{\rmn{CE}}$, the
      average progenitor primary mass, $M_{1}$, and the orbital
      period, $P_{\rmn{CE,i}}$, at the start of the CE phase and the
      average CE ejection efficiency, $\alpha_{\rmn{CE}}$, calculated
      with $\lambda_{\rmn{g}}$.}
    \begin{tabular}{@{}llllllllll@{}}
      \hline
      System & $M_{\rmn{WD}}/$M$_{\odot}$ &  $M_{2}/$M$_{\odot}$  &  $P_{\rmn{orb}}/$d & $P_{\rmn{CE}}/$d & $t_{\rmn{cool}}/$Gyr & $M_{1}/$M$_{\odot}$  &
      log($P_{\rmn{CE,i}}$/d) &  $\alpha_{\rmn{CE}}$  &  Ref.  \\
      \hline    
    SDSS1435+3733    &  0.505$\pm$0.025  &   0.218$\pm$0.028 &   0.126   &  0.134    & 0.275  &    1.26$\pm$0.30 &   2.81$\pm$0.14   &  0.26$\pm$0.18   & 1  \\       
    SDSS0052-0053   &  1.220$\pm$0.370  &   0.320$\pm$0.060  &   0.114   &   0.148   & 0.421  &     5.18$\pm$1.02 &  3.13$\pm$0.12  &   0.46$\pm$0.13   & 2  \\
    SDSS2123+0024   & 0.310$\pm$0.100   &   0.200$\pm$0.080  &   0.149   &   0.149   & 0.000  &     1.41$\pm$0.30 &  1.39$\pm$0.50  &   4.57$\pm$4.57   & 3  \\ 
    SDSS1529+0020   & 0.400$\pm$0.040   &  0.260$\pm$0.040   &   0.165   &   0.171   & 0.300  &     1.32$\pm$0.25 &  2.21$\pm$0.18  &   0.87$\pm$0.46   & 2  \\
    SDSS1411+1028   & 0.520$\pm$0.110   &  0.380$\pm$0.070   &   0.167   &  0.178    & 0.009  &     1.42$\pm$0.42 &  2.69$\pm$0.22  &  0.32$\pm$0.22    & 3 \\
    SDSS1548+4057   & 0.646$\pm$0.032   & 0.174$\pm$0.027    &   0.185   &   0.191   & 0.416  &     2.46$\pm$0.26 &  2.84$\pm$0.16  &  0.95$\pm$0.33    & 1 \\
    SDSS0303-0054   & 0.912$\pm$0.034   & 0.253$\pm$0.029   &    0.134   &  0.208    & $\ga{2.24}$  &     4.27$\pm$0.18 &  3.08$\pm$0.10  &     0.65$\pm$0.13 & 1\\
    SDSS2216+0102   & 0.400$\pm$0.060   & 0.200$\pm$0.080   &    0.210   & 0.213     & 0.297  &     1.32$\pm$0.25 &  2.18$\pm$0.26  &     1.29$\pm$0.77 & 3 \\ 
    SDSS1348+1834   & 0.590$\pm$0.040   & 0.319$\pm$0.060   &    0.249   & 0.252     & 0.184  &     2.03$\pm$0.37 &  2.83$\pm$0.20  &     0.66$\pm$0.30 & 3 \\
    SDSS0238-0005   &  0.590$\pm$0.220  &  0.380$\pm$0.070  &    0.212   & 0.239     & 0.045  &     1.49$\pm$0.50 &  2.58$\pm$0.31  &     0.49$\pm$0.35 & 3 \\
    SDSS2240-0935   & 0.410$\pm$0.080   & 0.250$\pm$0.120   &    0.261   & 0.265     & 0.443  &     1.32$\pm$0.25 &  2.19$\pm$0.31  &     1.20$\pm$0.83 & 3  \\
    SDSS1724+5620   & 0.420$\pm$0.010   & 0.360$\pm$0.070   &    0.333   & 0.333     & 0.000  &     1.30$\pm$0.25 &  2.39$\pm$0.09  &     0.81$\pm$0.41 & 2 \\
    SDSS2132+0031   & 0.380$\pm$0.040   & 0.320$\pm$0.010   &    0.222   & 0.224     & 0.179  &    1.34$\pm$0.25  &  2.08$\pm$0.19  &     1.10$\pm$0.56 & 3 \\ 
    SDSS0110+1326   & 0.470$\pm$0.020   & 0.310$\pm$0.050   &    0.333   & 0.333     & 0.051  &     1.21$\pm$0.23 &  2.67$\pm$0.10  &     0.48$\pm$0.25 & 1 \\
    SDSS1212-0123   & 0.470$\pm$0.010   & 0.280$\pm$0.020   &    0.333   & 0.334     & 0.191  &     1.19$\pm$0.22 &  2.70$\pm$0.09  &     0.48$\pm$0.25 & 4 \\
    SDSS1731+6233   & 0.450$\pm$0.080   & 0.320$\pm$0.010   &    0.268   & 0.271     & 0.228  &     1.28 $\pm$0.26 &  2.48$\pm$0.29 &      0.62$\pm$0.47& 3  \\
    SDSS1047+0523   & 0.380$\pm$0.200   & 0.260$\pm$0.040   &    0.382   & 0.384     & 0.417  &     1.38$\pm$0.31  &  1.74$\pm$0.71 &      4.68$\pm$4.68 & 5 \\
    SDSS1143+0009   & 0.620$\pm$0.070   & 0.320$\pm$0.010    &   0.386   & 0.387     & 0.138  &     2.14$\pm$0.41   &  2.84$\pm$0.19 &     0.87$\pm$0.38 & 3 \\       
    SDSS2114-0103   & 0.710$\pm$0.100   & 0.380$\pm$0.070   &    0.411   & 0.411     & 0.018  &    2.65$\pm$0.38   &  2.95$\pm$0.16 &      0.72$\pm$0.23 & 3 \\
    SDSS2120-0058   & 0.610$\pm$0.060   & 0.320$\pm$0.010  &     0.449   & 0.450     & 0.156  &    2.11$\pm$0.40   &  2.84$\pm$0.19 &      0.98$\pm$0.43 & 3 \\ 
    SDSS1429+5759   & 1.040$\pm$0.170   & 0.380$\pm$0.060  &     0.545   & 0.564     & 0.401  &    4.91$\pm$0.65   &   3.13$\pm$0.11 &      0.91$\pm$0.21 & 3 \\
    SDSS1524+5040   & 0.710$\pm$0.070   & 0.380$\pm$0.060  &     0.590   & 0.596     & 0.109  &    2.74$\pm$0.30   &  2.97$\pm$0.13 &    0.91$\pm$0.25 & 3 \\
    SDSS2339-0020   & 0.840$\pm$0.360  &  0.320$\pm$0.060  &     0.655   & 0.657     & 0.508  &     1.97$\pm$0.97  &   2.86$\pm$0.18  &     0.87$\pm$0.58 & 2 \\
    SDSS1558+2642   & 1.070$\pm$0.260  & 0.319 $\pm$0.060  &     0.662   & 0.665     & 0.609  &     4.77$\pm$0.95  &   3.12$\pm$0.11  &     1.17$\pm$0.30 & 3 \\
    SDSS1718+6101   & 0.520$\pm$0.090  & 0.320$\pm$0.010   &     0.673   & 0.673     & 0.075  &     1.44$\pm$0.42  &   2.73$\pm$0.19  &     0.85$\pm$0.57 & 3 \\             
    SDSS1414-0132   & 0.730$\pm$0.200  &  0.260$\pm$0.040  &     0.728   & 0.729     & 0.329  &   2.25$\pm$0.68    &    2.84$\pm$0.20 &      1.55$\pm$0.68 & 5 \\
    SDSS0246+0041   & 0.900$\pm$0.150  &  0.380$\pm$0.010  &     0.728   & 0.737     & 0.309  &   3.98$\pm$0.62    &   3.10$\pm$0.10  &      1.00$\pm$0.21 & 2 \\
    SDSS1705+2109   & 0.520$\pm$0.050  & 0.250$\pm$0.120   &     0.815   & 0.815     & 0.023  &    1.40$\pm$0.40   &     2.81$\pm$0.16 &      0.95$\pm$0.73 & 3 \\
    SDSS1506-0120   & 0.430$\pm$0.130  & 0.320$\pm$0.010   &     1.051   & 1.051     & 0.251  &   1.36$\pm$0.30    &    2.21 $\pm$0.47  &     2.82$\pm$2.82 & 3 \\
    SDSS1519+3536   & 0.560$\pm$0.040  &  0.200$\pm$0.080  &     1.567   & 1.567     & 0.065  &    1.78$\pm$0.35   &     2.74$\pm$0.18  &     3.24$\pm$1.64 & 3 \\
    SDSS1646+4223   & 0.550$\pm$0.090  &  0.250$\pm$0.120  &     1.595   & 1.595     & 0.093  &    1.55$\pm$0.48   &     2.78$\pm$0.17  &     1.86$\pm$1.33 & 3\\
    SDSS0924+0024   & 0.520$\pm$0.050  &  0.320$\pm$0.010  &     2.404   & 2.404     & 0.059  &    1.44$\pm$0.40   &     2.81$\pm$0.16  &     1.78$\pm$1.31 & 3 \\ 
    SDSS2318-0935   & 0.490$\pm$0.060  &  0.380$\pm$0.070  &     2.534   & 2.534     & 0.026  &    1.31$\pm$0.32   &     2.73$\pm$0.18  &     1.66$\pm$1.15 & 3 \\
    SDSS1434+5335   & 0.490$\pm$0.030  &  0.320$\pm$0.010  &     4.357   & 4.357     & 0.030  &   1.14$\pm$0.22    &    2.82$\pm$0.13   &    1.78$\pm$1.11  & 3 \\
      \hline
    \end{tabular}
    \label{CERecon_SDSS}

    (1)\citet{Pyrzas2009}; (2)\citet{Mansergas2008};
    (3)Nebot-G\'{o}mez-Mor\'{a}n et al. (2011, submitted);
    (4)\citet{Moran2009}; (5)\citet{Schreiber2008}

  \end{minipage}
\end{table*}

\begin{table*}
  \centering
  \begin{minipage}{170mm}
    \caption{The same as table \ref{CERecon_SDSS} but for the observed
      sample of PCEBs from the \citet{Ritter2003} catalogue, Edition 7.14
      (2010)}
    \begin{tabular}{@{}llllllllll@{}}
      \hline
      System & $M_{\rmn{WD}}/$M$_{\odot}$ &  $M_{2}/$M$_{\odot}$  &  $P_{\rmn{orb}}/$d & $P_{\rmn{CE}}/$d & $t_{\rmn{cool}}/$Gyr & $M_{1}/$M$_{\odot}$  &
      log($P_{\rmn{CE,i}}$/d) &  $\alpha_{\rmn{CE}}$  &  Ref.  \\
      \hline
       0137-3457      &   0.390$\pm$0.035 &  0.053$\pm$0.006  & 0.080    &   0.084     & 0.179  &    1.33$\pm$0.25   &     2.04$\pm$0.17  &     2.63$\pm$1.33 & 1\\ 
       HR Cam         &   0.410$\pm$0.010  & 0.096$\pm$0.004  & 0.103    &   0.104     & 0.118  &    1.32$\pm$0.25   &     2.23$\pm$0.09  &     1.41$\pm$0.72  & 2\\
       RR Cae         &   0.440$\pm$0.023  & 0.180$\pm$0.010  & 0.304    &   0.309     & 2.037  &    1.27$\pm$0.24   &     2.47$\pm$0.13  &     1.07$\pm$0.59  & 3,4 \\
       DE CVn         &   0.540$\pm$0.040  &  0.410$\pm$0.050 & 0.364    &   0.486     & 0.895  &    1.48$\pm$0.42   &     2.83$\pm$0.16  &     0.49$\pm$0.36  & 4,5\\ 
       J2130+4710     &   0.554$\pm$0.017  &  0.555$\pm$0.023 & 0.521    &   0.527     & 0.088  &    1.79$\pm$0.37   &     2.84$\pm$0.19  &     0.62$\pm$0.33  & 4,6 \\
       EG UMa         &   0.630$\pm$0.050  &  0.360$\pm$0.040 & 0.668    &   0.689     & 0.313  &    2.28$\pm$0.35   &     2.88$\pm$0.19  &     1.17$\pm$0.49  & 7,8\\  
       1857+5144      &   0.610$\pm$0.040  &  0.410$\pm$0.030 & 0.266    &   0.266     & 0.000  &    2.16$\pm$0.36   &     2.87$\pm$0.20  &     0.55$\pm$0.24  & 9,10\\
       BPM 71214      &   0.770$\pm$0.060  &  0.540 :         & 0.202    &   0.289     & 0.120  &    3.17$\pm$0.27   &     3.07$\pm$0.10  &     0.38$\pm$0.08  & 11,12,13\\
       QS Vir         &   0.780$\pm$0.040  &  0.430$\pm$0.040 & 0.151    &   0.315     & 0.370  &   3.26$\pm$0.19    &     3.07$\pm$0.10  &     0.48$\pm$0.10  & 4,14\\
       V471 Tau       &   0.840$\pm$0.050  &  0.930$\pm$0.070 & 0.521    &   0.521     & 0.001  &   3.70$\pm$0.25    &     3.15$\pm$0.10  &     0.35$\pm$0.07   & 15,16\\
       LM Com         &   0.350$\pm$0.030  &  0.170$\pm$0.020  & 0.259   &     0.260   & 0.032    &   1.36$\pm$0.26    &     1.82$\pm$0.16  &     2.88$\pm$1.33  & 17 \\ 
       MS Peg         &   0.490$\pm$0.040  &  0.190$\pm$0.020  & 0.174  &     0.174   & 0.027    &   1.25$\pm$0.28    &     2.73$\pm$0.15  &     0.40$\pm$0.26  & 17 \\
       GK Vir         &   0.510$\pm$0.040  &  0.100 :          & 0.344  &     0.344   & 0.002    &   1.39$\pm$0.36    &     2.72$\pm$0.15  &     1.29$\pm$0.92  & 4,18 \\
       NN Ser         & 0.535$\pm$0.012  &  0.110$\pm$0.004  & 0.130   &     0.130    & 0.001   &   1.68$\pm$0.27    &     2.66$\pm$0.17  &     1.02$\pm$0.49  & 19 \\
       1042-6902      &  0.560$\pm$0.050   &  0.140$\pm$0.010  & 0.337  &    0.340    & 0.080  &   1.57$\pm$0.45    &     2.75$\pm$0.17  &     1.10$\pm$0.81  & 21\\
       J2013+4002     &  0.560$\pm$0.030   &  0.230$\pm$0.010  & 0.706   &    0.710   &  0.002    &   1.84$\pm$0.35    &     2.76$\pm$0.19  &     1.74$\pm$0.88  & 20,21\\
       FS Cet         &  0.570$\pm$0.030  &  0.390$\pm$0.020  &  4.230  &    4.230    &  0.001   &   1.90$\pm$0.37    &     2.82$\pm$0.20  &     3.80$\pm$2.01  & 20,22\\
       IN CMa         &  0.580$\pm$0.030  &  0.430$\pm$0.030  &  1.260  &    1.260    &  0.002   &   1.97$\pm$0.37   &     2.85$\pm$0.20  &     1.51$\pm$0.73  & 11,20\\
       BE UMa         & 0.590$\pm$0.070   &  0.250$\pm$0.080  &  2.291  &    2.291    &  0.000   &   1.88$\pm$0.41    &    2.77$\pm$0.19  &     3.55$\pm$1.80  & 23,24 \\
       UZ Sex         & 0.680$\pm$0.230   &  0.220$\pm$0.050  &  0.597  &    0.597    &  0.084   &   1.72$\pm$0.69   &     2.78$\pm$0.18   &    1.12$\pm$0.72   & 25,26 \\
       J1016-0520AB   & 0.610$\pm$0.060  &  0.150$\pm$0.020   &  0.789  &    0.790    &  0.002   &   2.13$\pm$0.39   &      2.78$\pm$0.19  &     2.88$\pm$1.26  &  27 \\
       2009+6216      & 0.620$\pm$0.020  &  0.189$\pm$0.004  &   0.741  &    0.741    &  0.020 &   2.30$\pm$0.30   &     2.82$\pm$0.19   &     2.24$\pm$0.93   & 28\\
       CC Cet         & 0.400$\pm$0.110  &  0.180$\pm$0.050  &   0.287  &    0.287    & 0.000    &   1.37$\pm$0.30   &     1.95$\pm$0.44   &     2.57$\pm$2.49  & 25,29\\
       HZ 9           & 0.510$\pm$0.100  &  0.280$\pm$0.040  &  0.564  &    0.564    & 0.086    &   1.42$\pm$0.40   &     2.66$\pm$0.22   &    0.93$\pm$0.64   & 30,31\\
       LTT 560        & 0.520$\pm$0.120  &  0.190$\pm$0.050  &  0.148  &    0.162    & 1.040    &   1.44$\pm$0.42   &     2.61$\pm$0.24   &    0.60$\pm$0.36   & 32\\
       IK Peg         & 1.190$\pm$0.050  &  1.170 :          &  21.722 &   21.722     & 0.027   &   6.10$\pm$0.24   &     3.28$\pm$0.12   &    3.98$\pm$1.19   & 33,34\\ 
      \hline
    \end{tabular}

    \label{CERecon_RKCat}

    (1)\citet{maxted06}; (2)\citet{maxted98}; (3)\citet{Maxted2007};
    (4)\citet{Parsons2010}; (5)\citet{besselaar07};
    (6)\citet{maxted04}; (7)\citet{Bleach2002};
    (8)\citet{Shimanskii2002}; (9)\citet{aungwerojwit07};
    (10)\citet{Shimansky2009}; (11)\citet{kawka02};
    (12)\citet{Kawka2003}; (13)\citet{Kawka2005};
    (14)\citet{odonoghue03}; (15)\citet{obrien01};
    (16)\citet{Kaminsky2007}; (17)\citet{shimansky03};
    (18)\citet{fulbright93};
    (19)\citet{catalan94};(20)\citet{kawka08}; (21)\citet{good05};
    (22)\citet{vt94}; (23)\citet{Shimanskii2008};
    (24)\citet{Raguzova2003}; (25)\citet{saffer93};
    (26)\citet{Bruch1999}; (27)\citet{vennes99};
    (28)\citet{morales05}; (29)\citet{Somers1996}; (30)\citet{lp81};
    (31)\citet{stauffer87}; (32)\citet{tappert07}; (33)\citet{lsb93};
    (34)\citet{Vennes1998}

  \end{minipage}
\end{table*}

\begin{table*}
  \centering
    \caption{Similar to table \ref{CERecon_SDSS} but now showing the
      values of $M_{1}$, $P_{\rmn{CE,i}}$ and $\alpha_{\rmn{CE}}$ for
      the SDSS PCEBs, where consider the internal energy of the
      primary progenitor envelope, i.e. we use
      $\lambda=\lambda_{\rmn{b}}$.}
    \begin{tabular}{@{}llll@{}}
      \hline
      System & $M_{1}/$M$_{\odot}$  & log($P_{\rmn{CE,i}}$/d) &  $\alpha_{\rmn{CE}}$ \\
      \hline
      SDSS1435+3733  &  1.26$\pm$0.30    &    2.81$\pm$0.14   &    0.17$\pm$0.05      \\ 
      SDSS0052-0053  &  5.35$\pm$1.24    &    3.01$\pm$0.09   &    0.03$\pm$0.02  \\ 
      SDSS2123+0024  &  1.41$\pm$0.30    &    1.39$\pm$0.50   &    2.04$\pm$2.04   \\        
      SDSS1529+0020  &   1.32$\pm$0.25   &     2.21$\pm$0.18  &     0.37$\pm$0.21\\
      SDSS1411+1028  &   1.42$\pm$0.42   &     2.69$\pm$0.22  &     0.14$\pm$0.07\\
      SDSS1548+4057  &   2.46$\pm$0.26   &     2.84$\pm$0.16  &     0.13$\pm$0.11 \\
      SDSS0303-0054  &  4.37$\pm$0.16    &        2.96$\pm$0.03    &       0.03$\pm$0.02 \\
      SDSS2216+0102  &   1.32$\pm$0.25   &         2.18$\pm$0.26   &        0.56$\pm$0.34  \\  
      SDSS1348+1834  &   2.03$\pm$0.37   &         2.83$\pm$0.20   &        0.14$\pm$0.10\\
      SDSS0238-0005  &   1.48$\pm$0.49   &         2.58$\pm$0.31   &        0.18$\pm$0.14\\
      SDSS2240-0935 &   1.32$\pm$0.25   &         2.19$\pm$0.31   &        0.54$\pm$0.35\\
      SDSS1724+5620 &   1.30$\pm$0.25   &         2.39$\pm$0.09   &        0.34$\pm$0.18\\
      SDSS2132+0031 &   1.34$\pm$0.25   &         2.08$\pm$0.19   &        0.48$\pm$0.24  \\ 
      SDSS0110+1326 &   1.21$\pm$0.23   &         2.67$\pm$0.10   &        0.25$\pm$0.09\\
      SDSS1212-0123 &   1.19$\pm$0.22   &         2.70$\pm$0.09   &        0.26$\pm$0.08 \\
      SDSS1731+6233 &   1.28$\pm$0.26   &         2.48$\pm$0.29   &        0.30$\pm$0.18\\
      SDSS1047+0523 &   1.38$\pm$0.31   &         1.74$\pm$0.71   &        2.08$\pm$2.08  \\
      SDSS1143+0009 &   2.14$\pm$0.41   &         2.84$\pm$0.19   &        0.17$\pm$0.13    \\   
      SDSS2114-0103 &   2.62$\pm$0.37   &         2.93$\pm$0.15   &        0.07$\pm$0.07 \\
      SDSS2120-0058 &   2.11$\pm$0.40   &         2.84$\pm$0.19   &        0.20$\pm$0.15 \\
      SDSS1429+5759 &   4.84$\pm$0.61   &         2.99$\pm$0.03   &        0.05$\pm$0.03\\
      SDSS1524+5040 &   2.73$\pm$0.30   &         2.96$\pm$0.12   &        0.07$\pm$0.07\\
      SDSS2339-0020 &   1.80$\pm$0.72   &         2.84$\pm$0.17   &        0.27$\pm$0.20\\
      SDSS1558+2642 &   4.50$\pm$0.83   &         2.98$\pm$0.03   &        0.06$\pm$0.04\\
      SDSS1718+6101 &   1.44$\pm$0.42   &         2.73$\pm$0.19   &        0.37$\pm$0.18  \\            
      SDSS1414-0132 &   2.15$\pm$0.58   &         2.82$\pm$0.19   &        0.30$\pm$0.28 \\
      SDSS0246+0041 &   3.87$\pm$0.60   &         2.99$\pm$0.05   &        0.05$\pm$0.03 \\
      SDSS1705+2109 &   1.45$\pm$0.40   &         2.79$\pm$0.16   &        0.51$\pm$0.20\\
      SDSS1506-0120 &   1.36$\pm$0.30   &         2.21$\pm$0.47   &        1.26$\pm$1.24 \\
      SDSS1519+3536 &   1.78$\pm$0.35   &         2.74$\pm$0.18   &        1.02$\pm$0.54 \\
      SDSS1646+4223 &   1.55$\pm$0.48   &         2.78$\pm$0.17   &        0.76$\pm$0.38\\
      SDSS0924+0024 &   1.44$\pm$0.40   &         2.81$\pm$0.16   &        0.85$\pm$0.33 \\
      SDSS2318-0935 &   1.31$\pm$0.32   &         2.73$\pm$0.18   &        0.85$\pm$0.35\\
      SDSS1434+5335 &   1.14$\pm$0.22   &         2.82$\pm$0.13   &        1.20$\pm$0.39\\
   \hline
   \label{CERecon_SDSS_B}
    \end{tabular}
\end{table*}

\begin{table*}
  \centering
  \caption{Same as table \ref{CERecon_SDSS_B} but now showing the
    values of $M_{1}$, $P_{\rmn{CE,i}}$ and $\alpha_{\rmn{CE}}$ for
    the PCEBs from the \citet{Ritter2003} catalogue, Edition 7.14 (2010).}
    \begin{tabular}{@{}llll@{}}
      \hline
      System & $M_{1}/$M$_{\odot}$  & log($P_{\rmn{CE,i}}$/d) &  $\alpha_{\rmn{CE}}$ \\
      \hline
     0137-3457  &  1.33$\pm$0.25  &            2.04$\pm$0.17 &          1.12$\pm$0.57 \\ 
      HR Cam    &  1.32$\pm$0.25  &           2.23$\pm$0.09  &         0.59$\pm$0.31 \\
      RR Cae    &  1.27$\pm$0.24  &           2.47$\pm$0.12  &         0.48$\pm$0.23 \\
      DE CVn    &  1.48$\pm$0.42  &           2.83$\pm$0.16  &         0.22$\pm$0.09   \\
     J2130+4710 & 1.79$\pm$0.37   &          2.84$\pm$0.19   &        0.18$\pm$0.09 \\
      EG UMa    &  2.28$\pm$0.35  &           2.88$\pm$0.19  &         0.19$\pm$0.16  \\
     1857+5144 &  2.16$\pm$0.36  &            2.87$\pm$0.20 &        0.10$\pm$0.08  \\ 
     BPM 71214 &  3.17$\pm$0.26  &           3.02$\pm$0.08  &         0.018$\pm$0.017 \\  
     QS Vir    &  3.32$\pm$0.17  &           3.00$\pm$0.06  &         0.022$\pm$0.018 \\
      V471 Tau &  3.77$\pm$0.24  &           3.06$\pm$0.04  &         0.018$\pm$0.011 \\
     LM Com    &  1.36$\pm$0.26  &           1.82$\pm$0.16  &         1.26$\pm$0.58 \\
     MS Peg    &  1.25$\pm$0.28  &           2.73$\pm$0.15  &         0.23$\pm$0.07 \\
     GK Vir    &  1.39$\pm$0.36  &           2.72$\pm$0.15  &         0.68$\pm$0.23 \\
      NN Ser   &  1.68$\pm$0.27  &           2.66$\pm$0.17  &         0.37$\pm$0.14 \\
     1042-6902   &  1.57$\pm$0.45  &           2.75$\pm$0.17  &         0.46$\pm$0.21 \\
     J2013+4002  &  1.84$\pm$0.35  &           2.76$\pm$0.19  &         0.50$\pm$0.28  \\
     FS Cet      &  1.90$\pm$0.37  &           2.82$\pm$0.20  &         1.00$\pm$0.64 \\
     IN CMa      &  1.97$\pm$0.37  &           2.85$\pm$0.20  &         0.35$\pm$0.25 \\
    J1016-0520AB &  2.13$\pm$0.39  &           2.78$\pm$0.19  &         0.60$\pm$0.44 \\
    2009+6216    &  2.30$\pm$0.30  &           2.82$\pm$0.19  &         0.38$\pm$0.32\\
    CC Cet       &  1.37$\pm$0.30  &           1.95$\pm$0.44  &         1.15$\pm$1.11\\
    HZ 9         &  1.42$\pm$0.40  &           2.66$\pm$0.22  &         0.41$\pm$0.21\\
     LTT 560     &  1.44$\pm$0.42  &           2.61$\pm$0.24  &         0.25$\pm$0.14\\
     BE UMa      &  1.88$\pm$0.41  &           2.77$\pm$0.19  &         0.97$\pm$0.63\\
     UZ Sex      &  1.66$\pm$0.63  &           2.77$\pm$0.17  &         0.39$\pm$0.27\\
     IK Peg      &  6.29$\pm$0.17  &           3.10$\pm$0.01  &         0.19$\pm$0.06  \\
    \hline
    \end{tabular}
    \label{CERecon_RKCat_B}
\end{table*}

\end{document}